\documentclass[showpacs,amsmath,amssymb,endfloats*]{revtex4}
\usepackage{graphicx}
\newcommand{\LA}{\Lambda}\newcommand{\LAt}{\tilde{\Lambda}}
\newcommand{\beao}{\begin{eqnarray*}}\newcommand{\eeao}{\end{eqnarray*}}
\newcommand{\be}{\begin{equation}}\newcommand{\ee}{\end{equation}}
\newcommand{\bea}{\begin{eqnarray}}\newcommand{\eea}{\end{eqnarray}}
\newcommand{\beq}{\begin{eqnarray}}\newcommand{\eeq}{\end{eqnarray}}
\newcommand{\nn}{\nonumber}\newcommand{\pa}{\partial}
\newcommand{\ep}{\varepsilon}\newcommand{\om}{\omega}
\newcommand{\Om}{\Omega}\newcommand{\ga}{\gamma}
\newcommand{\al}{\alpha}
\newcommand{\Ref}[1]{(\ref{#1})}

\begin{document}
\thispagestyle{empty}
\title{First analytic correction beyond PFA for the electromagnetic field\\
in sphere-plane geometry}

\author{M.~Bordag\footnote{bordag@itp.uni-leipzig.de}}\author{V.~Nikolaev\footnote{Vladimir.Nikolaev@ide.hh.se}}

\affiliation{Leipzig University, Vor dem Hospitaltore 1, D-04103
Leipzig,
Germany  \\ and\\
Halmstad University, Box 823, S-30118 Halmstad}
\begin{abstract}
We consider the vacuum energy for a configuration of a sphere in
front of a plane, both obeying conductor boundary condition, at
small separation.  For the separation becoming small we derive the
first next-to-leading order of the asymptotic expansion in the
separation-to-radius ratio $\ep$. This correction is of order
$\ep$. In opposite to the scalar cases it contains also
contributions proportional to logarithms in first and second
order, $\ep \ln \ep$ and $\ep (\ln \ep)^2$. We compare this result
with the available findings of numerical and experimental
approaches.
\end{abstract}
 \maketitle

\section{Introduction}

The Proximity Force Approximation (PFA) is the most important
approximation for calculation of forces between curved surfaces at
small separation. It originates from a work on adhesion back in
1934 based on the simple idea to integrate the local force density
known from parallel surfaces \cite{derjaguin}. Given a
sufficiently fast  decrease of the force with separation this
method works independently from the kind of force. In this respect
it is universal. However, this method does not allow to beyond nor
does it give any information on its precision. Attempts to repeat
the original idea 'plane based' or 'sphere based', turned out to
give misleading results. It was only with the new method of
calculating the Casimir force, the T-matrix or TGTG-formula
approach that the door opened to go beyond PFA. In these
approaches, the interaction energy is represented by a infinite
dimensional determinant. At large and medium separation this
matrix can be truncated to become finite (even low) dimensional
and numerical evaluation is possible. At small separation this
does not work and below
\be\label{ep}
\ep\equiv\frac{d}{R}\sim 0.1
\ee
($d$-distance, $R$-radius of curvature, both at closest separation)
the numerical effort is unmanageable.

It must be mentioned that only $\ep\lesssim 0.1$ is the
experimentally interesting region for forces between macroscopic
bodies. This is because of the van der Waals and the Casimir
forces as being   quantum effects are generically microscopically
small and   become measurable only when multiplied by a
microscopically large interaction area. Since the forces decrease
proportional to $d^{-4}$ at large separation, only the combination
of small separation together with large radius of curvature allows
for measurements with appreciable precision. A typical value used
in experiments on precision measurements of the Casimir force
\cite{Harris2000} is $\ep\sim 10^{-3}$.

The interest in corrections beyond PFA is triggered from both,
theoretical and experimental sides. The first follows from the
challenge to  improve a situation which lasted more than 60 years,
the second from the high precision of the contemporary force
measurements and from the accuracy one would like to achieve for
their comparison with the theoretical predictions. It must be
mentioned that this has implications far beyond the atomic or
solid state physics as mean to obtain stronger constraints on new
physics ({\it Fifth Forces}), see for example \cite{Chen:2005as}.

For scalar fields, analytical corrections    beyond PFA  were
obtained as an asymptotic expansion
\begin{equation} \label{1.ep}
\frac{E}{E_{\rm PFA}}=1+\al\, \ep+\dots
\end{equation}
of the energy for $\ep\to0$ and simple  numbers were obtained for
the coefficient $\al$. In  \cite{Bordag:2006vc} this was done for
the geometry of a cylinder in front of a plane for a scalar field
obeying Dirichlet or Neumann boundary conditions. This includes
the electromagnetic field at once since its polarizations separate
in cylindrical geometry. The corrections for  a sphere in front of
a plane were obtained in \cite{BORDAG2008C} for a scalar field,
again for both, Dirichlet and Neumann boundary conditions.
However, in this case the corrections for the electromagnetic
field do not follow because of the non-separation of the
polarizations.

It is the aim of the present paper to fill this gap and to obtain
the first beyond-PFA corrections for the electromagnetic field
obeying conductor boundary conditions in the geometry of a sphere
in front of a  plane. Surprisingly, the asymptotic expansion in
this case turns out to contain logarithms, i.e., it has the form
\begin{equation} \label{eplnep}
\frac{E}{E_{\rm PFA}}=1+ \left(\al+\beta\,
\ln\ep+\gamma\, {(\ln\ep)}^2\right)\, \ep+\dots\,.
\end{equation}
The coefficients are calculated below and take the values
 $\al=-5.2$, $\beta=-0.0044$ and $\gamma=8.5~10^{-6}$. The logarithmic
 terms come in from contributions which are specific for the
 vector case.%, especially from the mixing of the polarizations.

In \cite{Decca2007} an experimental effort was undertaken to
measure the corrections beyond PFA by using several spheres whose
radii varied from 10 to 150 $\mu m$ at separations $d=200\dots
800\, nm$. The expansion  was assumed to have the form of
\Ref{1.ep} and the coefficient $\al$ was found to be zero within
the experimental precision. Also numerical efforts are reported
(for a cylinder in front of a plane in \cite{Lombardo:2008ww} and
for a sphere in \cite{Emig:2008zz,MaiaNeto:2008zz}) by pushing the
truncation in the T-matrix approach to higher orders and
extrapolating towards the known value at zero separation. The
results show agreement with the analytical results for Dirichlet
boundary conditions but not for Neumann boundary conditions;
details will be discussed in the last section. For a scalar field
obeying Dirichlet boundary conditions results were obtained  in
\cite{Gies:2006cq} using the independent method of world line
approach. Like the extrapolation these  confirm the analytical
results.

It is a second aim of the present paper to discuss in detail the
analytical corrections beyond PFA for all combinations of boundary
conditions for the scalar field. For instance, it will become
evident that and why the corrections for a sphere in front of a
plane are the same with Dirichlet boundary conditions on the
sphere, but Dirichlet or Neumann boundary conditions on the plane.
This case is interesting since the numerical results are different
for the two cases.

The paper is organized as follows. In the next section we give a
representation of the vacuum energy in the sphere-plane geometry
for all boundary conditions with special emphasis on the
translation formulas used. In the third section we re-derive the
asymptotic expansion for the scalar field and in the fourth
section we derive the expansion for the electromagnetic field. In
the last section we discuss the results. \\
Throughout the paper we use units with $\hbar=c=1$

%%%%%%%%%%%%%%%%%%%%%%%%%%%%%%%%%%%%%%%%%%%%%%%%%%%%%%%%%%%%%%%%%%%%%%%%%%
\section{Representation of the vacuum energy in sphere-plane
geometry}
The representation of the vacuum energy in sphere-plane geometry
was first derived in \cite{WIRZBA2006} for the scalar case within
the multiple scattering approach. Subsequently there appeared
numerous variations of the derivation; we use that in \cite{Buch},
~Chap. 10. Thereby we highlight one essential step - the use of
the translation formulas - having in mind their importance for
understanding the logarithmic contributions in the electromagnetic
case.

The general structure of the T-matrix representation of the vacuum
interaction  is
\be\label{2.E1}
E=\frac{1}{2}\int_{-\infty}^\infty\frac{d \xi}{2\pi}\,
{\rm Tr}\ln\left(1-GT\right).
\ee
The geometry is shown in Fig.1.

\includegraphics[width=7cm]{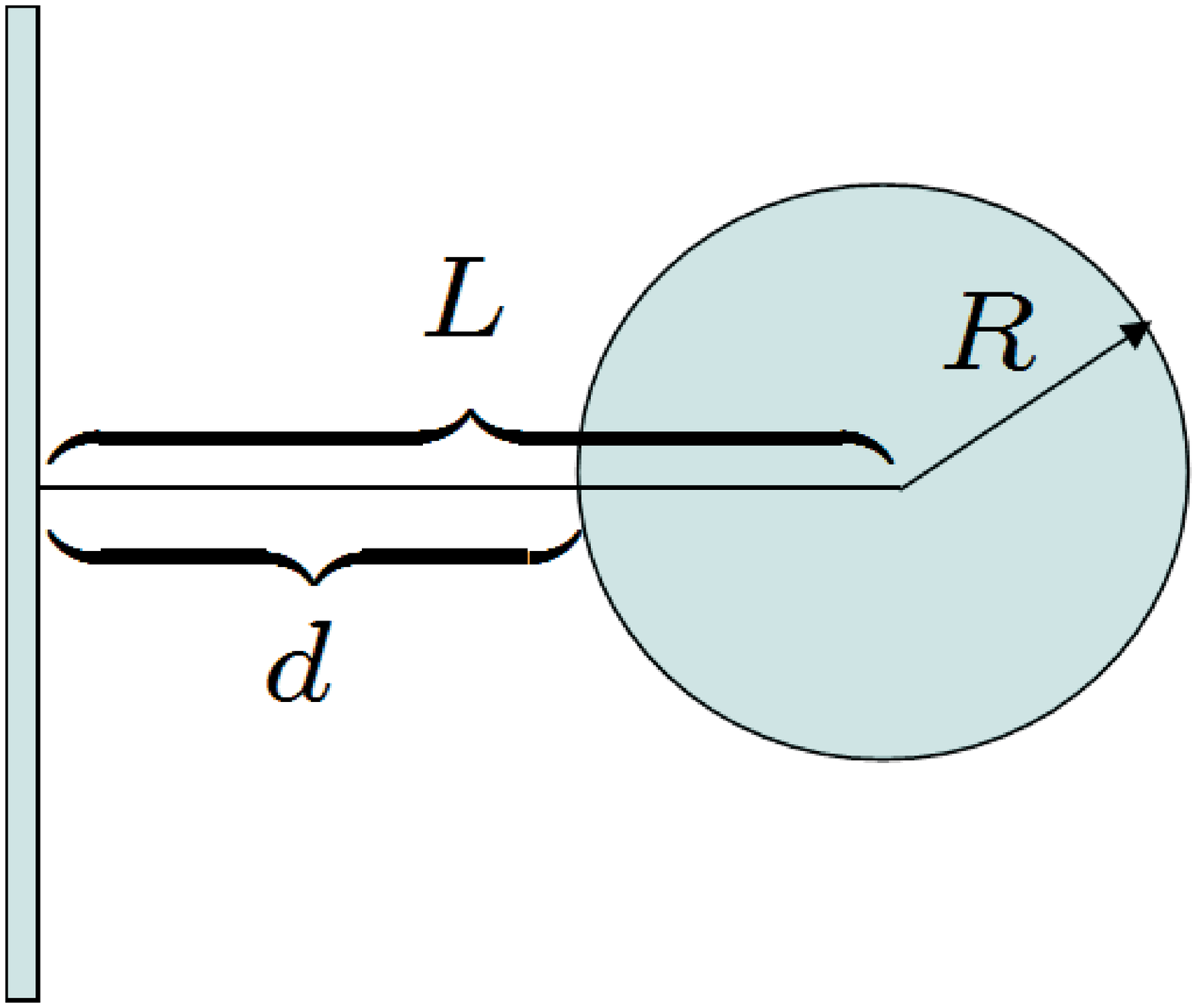}\\[0pt]
Fig.1: The configuration of a sphere in front of a plane.\\[20pt]

%\begin{figure}%\begin{picture}
%%\centering
% % \includegraphics[bb=  80 00 500 306,width=.6\textwidth]{Fig.eps}\\
% \includegraphics[width=7cm]{fig1.eps}\\[40pt]
%  \caption{The configuration of a sphere in front of a plane}\label{fig:1}
%\end{figure}
In \Ref{2.E1}
the frequency was rotated towards the imaginary axis, $\om\to i
\xi$. The symbol $G$ denotes the Greens function (or, strictly
speaking, the corresponding operator) describing the propagation
from the sphere to the mirror and back and $T$ is the T-matrix
operator for the scattering on the sphere. Using the basis
\be\label{2.basis}
u_{k,lm}(\mathbf{r})=j_l(kr)\,Y_{lm}(\Om_r),
 \ee
where the $Y_{lm}(\Om_r)$ are the spherical harmonics and $j(l(r)$
are the spherical Bessel functions,  the Greens function can be
written as
\be\label{2.Green}
G_\xi(\mathbf{r},\mathbf{r'})
=\frac{2}{\pi}\int_0^\infty\frac{dk\,k}{\xi^2+k^2}
\sum_{lm}u_{k,lm}(\mathbf{r})u^*_{k,lm}(\mathbf{r'}).
\ee
The limits of the summations are $l\ge 0$ and $|m|\le l$.

In \Ref{2.Green} both spatial arguments, $ \mathbf{r}$ and
$\mathbf{r'} $, appear to be in one and the same coordinate system
with spherical coordinates $(r,\Om_r)$. However, in the considered
geometry, we would like to expand the T-matrix operator in its own
coordinate system centered in $(0,0,d+R)$. If this system is taken
for $\mathbf{r}$, we need for $\mathbf{r'}$ a translation from
this one to the mirror and back. This can be achieved by the {\it
translation formula}
\be\label{2.transl}
u_{k,lm}(\mathbf{r}+a\mathbf{e}_z)
=\sum_{l'm'}A_{lm,l'm'}(a)\,u_{k,l'm'}(\mathbf{r})
 \ee
with $a=2L$  and $\mathbf{e}_z$ is the unit vector along the
$z$-axis.
 In \Ref{2.transl}, $A_{lm,l'm'}(a)$ are the
translation coefficients. These involve the Clebsch-Gordan
coefficients, for details and an explicate expression see, for
example, Eq.(10.125) in \cite{Buch}. Applying these formulas,
after some transformations, the energy \Ref{2.E1} takes the form
\be\label{2.E2}
E=\frac{1}{2}\int_{-\infty}^\infty\frac{d\xi}{2\pi}\, {\rm
Tr}\ln\left(1-N\right),
 \ee
where the trace is the orbital momentum sum,
\be\label{2.tr1} {\rm Tr}=\sum_{m=0}^\infty\,
\sum_{l=|m|}^\infty\,.
 \ee
In \Ref{2.E2}, $N$ is an infinite dimensional  matrix in the
orbital momentum index, $N_{l,l'}$. Using the known relation ${\rm
Tr}\ln=\ln\det$ the energy can  represented as a determinant.
However, we do not use this in the following.

The entries of the matrix $N$ are, for Dirichlet boundary
conditions on the sphere and with the notation $N^{\rm D}_{l,l'}$
for $N_{l,l'}$,
\be\label{2.ND}
N^{\rm D}_{l,l'}
=\sqrt{\frac{\pi}{4\xi
L}}\sum_{l''=|l-l'|}^{l+l'}K_{l''+1/2}(2\xi L)H_{ll'}^{l''}\,
d^{\rm D}_l(\xi R) \,.
\ee
The function
\be\label{2.dD}
d^{\rm D}_l(\xi R)=\frac{I_{l+1/2}(\xi R)}{K_{l+1/2}(\xi R)}
\ee
is up to a factor the T-matrix for the scattering of a scalar field on
a hard sphere in orbital momentum representation. The corresponding expression
$N^{\rm N}_{l,l'}$ for Neumann boundary conditions on the sphere can be
obtained with
\be\label{2.dN} d^{\rm N}_l(\xi R)=\frac{\left(I_{l+1/2}(\xi
R)/\sqrt{\xi R}\right)'}{\left(K_{l+1/2}(\xi R)/\sqrt{\xi R}\right)'}
\ee
in place of $d^{\rm D}_l(\xi R)$ in \Ref{2.ND}. In the above
formulas $I_\nu$ and $K_\nu$ are the modified Bessel functions. We
note that in opposite to eqn. (10.140) in \cite{Buch} the function
$I_{l+1/2}(\xi R)$ carries the index $l$ in place of $l'$; a
substitution which is allowed under the trace in \Ref{2.E2}. In
\Ref{2.ND} we used the notation
\bea\label{2.H} %\hspace*{-112pt}
H_{ll'}^{l''}&=&  \sqrt{(2l+1)(2l'+1)}(2l''+1) \nn\\&&\times
\left(\begin{array}{ccc}l&l'&l''\\0&0&0\end{array}\right)
\left(\begin{array}{ccc}l&l'&l''\\m&-m&0\end{array}\right), \eea
which involves the Clebsch-Gordan coefficients coming in from the
translation formula. We use the notation of the $3j$-symbols.

Eq. \Ref{2.E2} represents the energy with Dirichlet boundary
conditions on the plane. The case of Neumann boundary conditions
on the plane is obtained by changing the signs in front of $N$. We
unite all four combinations of boundary condition in
\be\label{2.E4} E^{\rm XY}
=\frac{1}{2}\int_{-\infty}^\infty\frac{d\xi}{2\pi}\, {\rm
Tr}\ln\left(1-(-1)^x N^{\rm Y}\right). \ee
The first index denotes the boundary conditions on the plane, $\rm
X=D$ with $x=+1$ for Dirichlet and $\rm X=N$ with $x=-1$ for
Neumann conditions. The second index denotes the boundary
conditions on the sphere with $\rm Y=D$ for Dirichlet and $\rm
Y=N$ for Neumann conditions.
In the following we write all formulas
for Dirichlet boundary conditions on both and discuss the other
cases at the end of the next section.

The logarithm in \Ref{2.E2} is taken of a matrix. In the following
we will use the expansion of this logarithm such that the formula
for the energy takes the form
\bea\label{2.E3} E^{\rm }
&=&\frac{1}{2}\int_{-\infty}^\infty\frac{d\xi}{2\pi}\,
\sum_{s=0}^\infty\frac{-1}{s+1}
\\\nn&&   \times
\sum_{m=0}^\infty\, \sum_{l=|m|}^\infty\, \left(\prod_{j=1}^s
\sum_{l_j=|m|}^\infty\right) \,  \left( \prod_{i=0}^sN^{\rm }
_{l_i,l_{i+1}}\right)
 \eea
with the formal setting $l_0=l_{s+1}=l$. It should be mentioned
that the $N_{l,l'}$ are  diagonal in the azimutal index $m$ such
that the sum over $m$ appears only once.

The expansion \Ref{2.E3} corresponds to a perturbation series with
respect to the T-matrix of the scattering on the sphere. This
series appears to converge. The convergence can be easily seen for
large separations where it corresponds to the multipole expansion
and also for short separations as we will see below. It is known
that the vacuum energy for a transparent sphere, or for a
background potential, where it makes sense to introduce a coupling
constant, has the same structure as Eq.\Ref{2.E2}. The last
equation,
 in this way, appears
as a perturbative expansion with respect to the coupling constant.
Examples for the first order of this expansion  are considered in
a number of papers, see for example
\cite{Milton:2008vr,Milton:2009gk}.

The T-matrix representation for the electromagnetic field has the
same general structure as that for the scalar field. The main
difference is in the presence of the polarizations. The expansion
basis for the electromagnetic field has two components,
\bea\label{2.basis2} \mathbf{m}_{k,lm}(\mathbf{r}) &\equiv&
\mathbf{u}_{k,1lm}(\mathbf{r}) =\mathbf{L}
\frac{1}{\sqrt{\mathbf{L}^2}}\,u_{k,lm}(\mathbf{r})\,,
\\\nn
\mathbf{n}_{k,lm}(\mathbf{r}) &\equiv&
\mathbf{u}_{k,2lm}(\mathbf{r}) =\mathbf{\nabla}\times\mathbf{L}
\frac{1}{\sqrt{-\Delta}}\frac{1}{\sqrt{\mathbf{L}^2}}\,u_{k,lm}(\mathbf{r})\,,
\eea
where $\mathbf{L}$ is the orbital momentum operator. Under a
translation these functions mix and in place of \Ref{2.transl} the
translation formula is now
\bea\label{2.trans2}
&&\mathbf{m}_{k,lm}(\mathbf{r}+a\mathbf{e}_z)
\\\nn&&=
\sum_{l'm'}\left( B_{lm,l'm'}(a)\, \mathbf{m}_{k,l'm'}(\mathbf{r})
+ C_{lm,l'm'}(a)\, \mathbf{n}_{k,l'm'}(\mathbf{r})    \right),
\nn\\
&&\mathbf{n}_{k,lm}(\mathbf{r}+a\mathbf{e}_z)=
\nn\\\nn&&=\sum_{l'm'}\left( C_{lm,l'm'}(a)\,
\mathbf{m}_{k,l'm'}(\mathbf{r}) + B_{lm,l'm'}(a)\,
\mathbf{n}_{k,l'm'}(\mathbf{r})    \right). \eea
The Greens function (now in fact the Greens dyadic) can be
expressed in the basis \Ref{2.basis2} similar to \Ref{2.Green} and
reads
\be\label{2.Green2}
\mathbf{G}_\xi(\mathbf{r},\mathbf{r'})
=\frac{2}{\pi}\int_0^\infty\frac{dk\,k}{\xi^2+k^2}
\sum_{slm}\mathbf{u}_{k,slm}(\mathbf{r})\mathbf{u}^*_{k,slm}(\mathbf{r'}),
\ee
$s$ taking values $s=1,2$ and the s-wave is excluded, $l\ge 1$.
The translation coefficients $B_{lm,l'm'}(a)$ and $C_{lm,l'm'}(a)$
are known in the electromagnetic theory, as pointed out in
\cite{Emig:2007cf} a particularly useful representation can be
found in \cite{Wittmann1988}. In the given geometry, the
coefficient $B_{lm,l'm'}$
 can be
obtained by the substitution
\bea\label{2.BC}
H_{ll'}^{l''} \, \to \, H_{ll'}^{l''}\Lambda_{ll'}^{l''}
%B_{lm,l'm'}&=&\Lambda_{ll'}^{l''}\,A_{lm,l'm'}\,,
\eea
in $A_{lm,l'm'}$ with
\be\label{2.LA}
\Lambda_{ll'}^{l''}=\frac{\frac12\left[l''(l''+1)-l(l+1)-l'(l'+1)\right]}
{\sqrt{l(l+1)l'(l'+1)}}\,.
\ee
%.
The other one is given by
\bea\label{2.C}
C_{lm,l'm'}&=&\tilde{\Lambda}_{ll'}\,A_{lm,l'm'}\, \eea
with
\bea\label{2.LAt} \tilde{\Lambda}_{ll'}&=&\frac{2m\xi (d+R)}
{\sqrt{l(l+1)l'(l'+1)}},
 \eea
where the $A_{lm,l'm'}$ are the same as in the scalar case,
Eq.\Ref{2.transl}. The coefficients
$\Lambda_{ll'}^{l''}$ and $\tilde{\Lambda}_{ll'}$
result from the orbital momentum operators, for instance from the
normalization factors in \Ref{2.basis2}.

The translation from one coordinate system to the other mixes the
polarizations of the electromagnetic field. Therefore the energy
has now the representation
\be\label{2.E5}
E=\frac{1}{2}\int_{-\infty}^\infty\frac{d\xi}{2\pi}\, {\rm
Tr}\ln\left(1-\mathbb{N}\right),
 \ee
and the trace is
\be\label{2.tr2} {\rm Tr}=\sum_{m=0}^\infty\,
\sum_{l=max(1,|m|)}^\infty\,{\rm tr}\,,
 \ee
where tr denotes the trace over $\mathbb{N}$ which is now a
(2x2)-matrix in the  polarizations,
\bea\label{2.NED} \mathbb{N}_{l,l'}&\equiv&
\left(\begin{array}{cc} N^{(11)}_{l,l'}& N^{(12)}_{l,l'} \nn\\
N^{(21)}_{l,l'}& N^{(22)}_{l,l'}\end{array}\right)
\\&=&
\sqrt{\frac{\pi}{4\xi L}}\sum_{l''=|l-l'|}^{l+l'}K_{l''+1/2}(2\xi L)
H_{ll'}^{l''}\,
\\&&\nn
\times\left(\begin{array}{cc}\Lambda_{l,l'}^{l''}&\tilde{\Lambda}_{l,l'}
\\ \tilde{\Lambda}_{l,l'}&\Lambda_{l,l'}^{l''}\end{array}\right)
\left(\begin{array}{cc}d^{\rm TE}_l(\xi R)&0
\\ 0&-d^{\rm TM}_l(\xi R)\end{array}\right)
 \,.
\eea
%
%This is also a matrix in the indexes $l$ and $l'$, like in the scalar case.
The functions $d^{\rm TE}_l$ and $d^{\rm TM}_l$ describe the
scattering of the corresponding polarizations of the
electromagnetic field on a conducting sphere and are similar to
that of the scalar field. For the TE mode it is literally the
same,
\be\label{2.dTE} d^{\rm TE}_l(\xi R)= \frac{I_{l+1/2}(\xi
R)}{K_{l+1/2}(\xi R)}\,, \ee
and for the TM-mode it is
\be\label{2.dTM} d^{\rm TM}_l(\xi R)= \frac{\left(I_{l+1/2}(\xi
R)\sqrt{\xi R}\right)'} {\left(K_{l+1/2}(\xi R)\sqrt{\xi
R}\right)'}\,. \ee
The minus sign in front of $d^{\rm TM}_l$ in \Ref{2.NED} results
from the   the spin of the electromagnetic field under reflection
on the plane.

Representation \Ref{2.E5} of the vacuum energy of the
electromagnetic field was derived in different notations in
\cite{Emig:2008zz,MaiaNeto:2008zz}. The coincidence with these
formulas can be checked by comparing some first orders of the
expansion for large separation.
%%%%%%%%%%%%%%%%%%%%%%%%%%%%%%%%%%%%%%%%%%%%%%%%%%%%%%%%%%%%%%%%%%%%%%%%%%
%%%%%%%%%%%%%%%%%%%%%%%%%%%%%%%%%%%%%%%%%%%%%%%%%%%%%%%%%%%%%%%%%%%%%%%%%%
%%%%%%%%%%%%%%%%%%%%%%%%%%%%%%%%%%%%%%%%%%%%%%%%%%%%%%%%%%%%%%%%%%%%%%%%%%
%%%%%%%%%%%%%%%%%%%%%%%%%%%%%%%%%%%%%%%%%%%%%%%%%%%%%%%%%%%%%%%%%%%%%%%%%%
\section{The asymptotic expansion for the scalar field}
In this section we consider the asymptotic expansion for the
scalar field at small separation. We follow \cite{BORDAG2008C} and
repeat the main steps of the derivation since these appear
essentially in the same form in the next section for the
electromagnetic case. At once we add a discussion on all
combinations of the boundary conditions on the plane and on the
sphere.

The vacuum energy is given by Eq.\Ref{2.E3}. First of all we make
the substitution $\xi\to\xi/R$ to get rid of the dimensional
variables.
 Then, as already mentioned, for
decreasing separation the convergence of the integral and the sums
in \Ref{2.E3} slows down and the main contribution comes from
higher and higher frequencies and orbital momenta. We know, by
hindsight, the region delivering the dominating contributions.
Since  at small separation all summation indices involved take high
values we substitute all sums by corresponding integrations. In
this way we drop exponentially small contributions which is
allowed aiming for an asymptotic expansion. In these integrations
we make the substitutions
\bea\label{3.subst} \xi=\frac{t}{\ep}\,\sqrt{1-\tau^2}, \quad
l=\frac{t}{\ep}\,\tau, \quad m=\sqrt{\frac{t\tau}{\ep}}\, \mu\,,
\nn \\
%&& l_1=\sqrt{\frac{4t}{\ep}}\,n_1, \dots , l_s=\sqrt{\frac{4t}{\ep}}\,n_s\,,
\quad  \tilde{l}_i=\sqrt{\frac{4t}{\ep}}\,n_i~~(i=1,\dots,s)\,,
\eea
where we divided the orbital momenta by means of
$l_i=l+\tilde{l}_i$ ($i=1,\dots,s$) into the index $l$ of the main
diagonal and the off-diagonal indices $\tilde{l}_i$. The variable
$\tau$ has the meaning of the cosine of the polar angle in the
$\xi,l$-plane. This substitution describes the region where the
main contributions come from.

In the new variables the expression for the energy reads
\bea\label{3.E1} E&=& -\frac{R}{4\pi d^2} \sum_{s=0}^\infty
\frac{1}{s+1} \int_0^\infty dt \, t  \ e^{-2t(s+1)} \nn\\&&\times
\int_{0}^1 \frac{d\tau\,\sqrt{\tau}}{\sqrt{1-\tau^2}}\,
\int_{-\infty}^\infty \frac{d\mu}{\sqrt{\pi}}\
e^{-\mu^2(s+1)/\tau}\ \nn\\&& \times \left(\prod_{j=1}^s
\int_{n_0}^\infty\frac{dn_j}{\sqrt{\pi}} \right)
  \cal{Z}^{\rm }\,,
% \nn \\ &&  ~~~~~~~~~~~~~~\times
%\int_{-\infty}^\infty \frac{dn_1}{\sqrt{\pi}}\, \dots
%\int_{-\infty}^\infty \frac{dn_s}{\sqrt{\pi}}\ {\cal M}  \
\eea
where
\be\label{3.n0}n_0=
-\frac{\tau}{2}\sqrt{\frac{t}{\ep}}+\frac12 |\mu|\sqrt{\tau}\,
\ee
is the lower boundary in the $n_j$-integrations. It
follows with \Ref{3.subst} from $l\ge |m|$. In \Ref{3.E1},
\be\label{3.Z} {\cal Z}^{\rm }=
    \prod_{i=0}^s   \left( \sqrt{\frac{4\pi t}{\ep}}\,
    N^{\rm }_{l+\tilde{l}_i,l+\tilde{l}_{i+1}}\right)\,e^{\eta_{\rm as}}\,,
 \ee
collects the information from the scattering process
together with the prefactors which follow from the substitution
\Ref{3.subst}. In \Ref{3.Z} we use the formal definitions
$\tilde{l}_0=\tilde{l}_{s+1}=0$ and we defined
\be\label{3.eta} \eta_{\rm as}=2t(s+1)+\eta_1+\mu^2 \,
\frac{s+1}{\tau}\ee
with $\eta_1=\sum_{i=0}^s(n_i-n_{i+1})^2$.

Next we expand ${\cal Z}^{\rm }$ for small $\ep$. It turns out
that it is possible to do this expansion straightforwardly by
expanding all quantities entering. These are the Bessel functions
and the Clebsch-Gordan coefficients in the $N_{l,l'}$,
Eq.\Ref{2.ND}. Here we can follow the corresponding expansion in
\cite{BORDAG2008C}. While for the Bessel functions the known
uniform asymptotic expansion can be used, for the Clebsch-Gordan
coefficients the corresponding asymptotic expansion was first
derived in \cite{BORDAG2008C}. It rests on an integral
representation of these coefficients. In this way, one arrives at
an asymptotic expansion, $N_{l,l'}\to N^{\rm as}_{l,l'}$, with
\bea\label{3.Nlls}
N^{\rm as}_{l,l'}&=&
\sqrt{\frac{\ep \tau}{2\pi
t(1+\tau)}}\ e^{-2t-(n-n')^2}
\int\limits_{-\infty}^{\infty}
\frac{d\eta}{\sqrt{\pi}}\, e^{-\eta^2+2i\eta\sqrt{2}\mu+\mu^2}	
\\&&\nn\times
\sum\limits_{\nu=0}^{\nu_m}
  \frac{\eta^{2\nu}}{\nu!}\, \left(\frac{1-\tau}{1+\tau}\right)^{\nu}
\left(1+\sqrt{\ep} f(\eta,\mu,t\tau)+\dots\right).
%\\
%&=&\sqrt{\frac{\ep}{4\pi t}}	\, e^{-\eta_{\rm as}}\left(1+a(n,n')\sqrt{\ep}+b(n,n')\ep+\dots\right)
\eea
The function $f(\eta,\mu,t\tau)$ and the corresponding function in
the next order can be found in \cite{BORDAG2008C}.  The
integration over $\eta$ results from the mentioned integral
representation and the sum over $\nu$ from the summation over
$l''$ in \Ref{2.ND} with the substitution $l''=l+l'-2\nu$. By the
symmetry properties of the $3j$-symbols in \Ref{2.H} it follows
that $\nu$ takes integer values only. The upper limit of the
summation over $\nu$ is
\be\label{3.num} \nu_m=\frac{1}{2}(l+l'-|l-l'|). \ee
The   exponential factor follows from the function $\eta(z)$ in
the asymptotic expansion of the Bessel functions.

By means of the substitution \Ref{3.subst}, the upper limit of the
$\nu$ summation depends on $\ep$,
\be\label{3.num1}
\nu_m=\frac{t\tau}{\ep}-\sqrt{\frac{4t}{\ep}}\,|n-n'|.
%\qquad n_0=-\tau\sqrt{\frac{t}{\ep}}.
\ee
In the considered case of a scalar field it is possible to put
$\ep=0$ here. After that the sum over $\nu$ in \Ref{3.Nlls} can be
carried out. The integral over $\eta$ is Gaussian and can be
carried out too. We get
\bea\label{3.Nlls1} &&N^{\rm as}_{l,l'}
\\\nn&&= \sqrt{\frac{\ep}{4\pi t}}	\ e^{-2t-(n-n')^2-\mu^2/\tau}
      \left(1+a^{(1/2)}_{n,n'}\,\sqrt{\ep}+a^{(1)}_{n,n'}\,\ep+\dots\right).
\eea
The functions $a^{(1/2)}(n,n')$ and $a^{(1)}(n,n')$   are given in
Eq.(A.22) in \cite{BORDAG2008C}.

Eq.\Ref{3.Nlls1} must be inserted into Eq.\Ref{3.Z}. The
prefactors and the exponentials just cancel and the remaining
dependence on $\ep$ is contained in the bracket.  This fact
justifies the substitution \Ref{3.subst}. When inserting $N^{\rm
as}_{l,l'}$ into \Ref{3.Z} we get the asymptotic expansion ${\cal
Z}^{\rm as}$. After a re-expansion it takes the form
\bea\label{3.Zas  }
&&{\cal Z}^{\rm as}=
1+\sum_{i=1}^s a^{(1/2)}_{n_i,n_{i+1}}\ \sqrt{\ep}
\\\nn&&
+a^{\rm D}\ \ep+\dots\,.
 \eea
with %
\be\label{3.ad}
a^{\rm D}= \sum_{0<i<j<s}a^{(1/2)}_{n_i,n_{i+1}}a^{(1/2)}_{n_j,n_{j+1}}
+\sum_{i=1}^s a^{(1)}_{n_i,n_{i+1}} \,.
\ee
We mention that we used in the re-expansion the general formula
\be\label{re-exp}
\prod_{i=0}^s\left(1+x_i\right)=1+\sum_{i=0}^s x_i+\sum_{0<i<j<s}x_ix_j+\dots\,,
\ee
which holds in the sense of an expansion for small $x_i$. We note
that the product turned into sums.  We will use this formula below
several times without further notice.

Inserting this expression for $\cal Z^{\rm as}$ into \Ref{3.E1} we
get the asymptotic expansion of the energy. Here we have still an
$\ep$-dependence in $n_0$, \Ref{}. However,   we are allowed to
put this $\ep=0$, i.e., to take $n_0=-\infty$, since in doing so
all integrations remain finite. After that the integrations can be
carried out. These are either over simple exponentials or are
Gaussian. We mention the most complicated one, which is that over
the $n_j$. It was calculated in \cite{Bordag:2006vc}, Eq.(66),
\be\label{66} \left(\prod_{j=1}^s
\int_{-\infty}^\infty\frac{dn_j}{\sqrt{\pi}}
\right)\,e^{-\eta_1}=\frac{1}{\sqrt{s+1}}. \ee
The result for the energy is
\be\label{3.E2} E=
-\frac{1}{16\pi}\frac{R}{d^2}
\sum_{s=0}^\infty\frac{1}{(s+1)^4}\left(1+\frac{1}{3}\ \ep+\dots\right).
\ee
Carrying out the summation over $s$, in the leading order just the PFA emerges.
In the
order $\sqrt{\ep}$ the integrations gave zero for symmetry reasons
and in order $\ep$ we get the first correction beyond PFA.

%The energy, Eq.\Ref{3.E2}, is for Dirichlet boundary conditions. The
%case of Neumann boundary conditions on the sphere can be handled
%in complete analogy. There are only two differences. The first one
%is an additional contribution resulting from the derivatives in
%\Ref{2.dN}. As shown in \cite{BORDAG2008C}, it results in   additional
%contributions to \Ref{3.Zas}. The second
%difference is a minus sign resulting from the derivative of the
%Bessel function $K_\nu$ in the denominator in \Ref{2.dN}. It
%appears in each factor $N_{l,l'}$, hence it gives a sign factor
%$(-1)^{s+1}$ to the sum over $s$. Now we restore the notation of
%Eq.\Ref{2.E4} for the different combination of the boundary
%conditions and get
The energy, Eq.\Ref{3.E2}, is for Dirichlet boundary conditions on
both surfaces. The case of Neumann boundary conditions on the
sphere can be handled in complete analogy. There are only two
differences. The first one is an additional  minus sign resulting
from the derivative of the Bessel function $K_\nu$ in the
denominator in \Ref{2.dN}. It appears in each factor $N_{l,l'}$,
hence it gives a sign factor $(-1)^{s+1}$ to the sum over $s$. At
this place we restore the notation of Eq.\Ref{2.E4} for the
different combination of the boundary conditions and accounting
for all signs we get
\bea\label{3.E3} E^{\rm XD}&=&\frac{1}{16\pi}\frac{R}{d^2}
\sum_{s=0}^\infty\frac{(-1)^{1+x(s-1)}}{(s+1)^4}\left(1+\frac{1}{3}\
\ep+\dots\right), \\\nn
 E^{\rm XN}&=&\frac{1}{16\pi}\frac{R}{d^2}
\sum_{s=0}^\infty\frac{(-1)^{x(s-1)+s}}{(s+1)^4}
    \\\nn&&\times       \left(1+\left(\frac{1}{3}-\frac{2}{3}(s+1)^2\right)
\ep+\dots\right).
\eea
The second difference for Neumann boundary conditions  on the
sphere is an additional contribution containing the  factor of
$(s+1)^2$ in the last line. It results from the change in the
Debye polynomials due to the derivatives in \Ref{2.dN} in the
asymptotic expansion of the Bessel functions and results in a
different function $a^{\rm N}$ in place of \Ref{3.ad}. As shown in
\cite{BORDAG2008C}, this factor is the same even if generalizing
to Robin boundary conditions. For instance, it is the same for the
TM mode of the electromagnetic field in the next section.

In Eq.\Ref{3.E3} the summations result in Riemann zeta functions;
we need
\bea\label{}
&&{\sum_{s=0}^\infty \frac{1}{(s+1)^2}= \zeta(2) =\frac{\pi^2}{6}}, \quad
{\sum_{s=0}^\infty \frac{(-1)^s}{(s+1)^2}=\frac{1}{2} \zeta(2)}, \quad
\nn\\\nn
&&{ \sum_{s=0}^\infty \frac{1}{(s+1)^4}=  \zeta(4)=\frac{\pi^4}{90}},\quad
{ \sum_{s=0}^\infty \frac{(-1)^s}{(s+1)^4}=\frac{7}{8} \zeta(4)}.
\eea
In each case the leading order in \Ref{3.E3} delivers the PFA. The
relative corrections are
\bea\label{3.E4}
\frac{E^{\rm DD}}{E_{\rm PFA}}&=&1+\frac{1}{3}\ \ep+\dots \,,\nn \\
\frac{E^{\rm NN}}{E_{\rm PFA}}&=&1+\left(\frac{1}{3}-\frac{10}{\pi^2}\right)\ \ep+\dots \,,\nn \\
\frac{E^{\rm DN}}{E_{\rm PFA}}&=&1+\left(\frac{1}{3}-\frac{5}{\pi^2}\right)\ \ep+\dots \nn \,,\\
\frac{E^{\rm ND}}{E_{\rm PFA}}&=&1+\frac{1}{3}\ \ep+\dots\,.
\eea
We note that the corrections for Dirichlet boundary conditions on
the sphere, but Dirichlet or Neumann boundary conditions on the
plane (first and last line) are the same. This follows simply from
the structure of the signs in \Ref{3.E3}.
%%%%%%%%%%%%%%%%%%%%%%%%%%%%%%%%%%%%%%%%%%%%%%%%%%%%%%%%%%%%%%%%%%%%%%%%%%%
%%%%%%%%%%%%%%%%%%%%%%%%%%%%%%%%%%%%%%%%%%%%%%%%%%%%%%%%%%%%%%%%%%%%%%%%%%%
%%%%%%%%%%%%%%%%%%%%%%%%%%%%%%%%%%%%%%%%%%%%%%%%%%%%%%%%%%%%%%%%%%%%%%%%%%%
\section{The asymptotic expansion for the electromagnetic field}
The asymptotic expansion for the electromagnetic field start with
the same steps as in the scalar case, i.e., with the substitution
of the orbital momentum sums by integrals and the substitution
\Ref{3.subst}. The next step is the asymptotic expansion of the
functions $N_{l,l'}^{(ss')}$ entering the matrix $\mathbb{N}_{l,l'}$ in
Eq.\Ref{2.NED}. In the functions $d_l^{\rm TE}$, Eq.\Ref{2.dTE},
and $d_l^{\rm TM}$, Eq.\Ref{2.dTM}, we use again the uniform
asymptotic expansion of the Bessel functions. In the factors
$\Lambda_{ll'}^{l''}$ and $\tilde{\Lambda}_{ll'}^{l''}$, Eq.
\Ref{2.LA}, we use directly the substitution \Ref{3.subst} and
write them in the form
\bea \label{4.LA}
 &&   \Lambda_{l,l'}^{l''}	\equiv 	1+\ep\,
\lambda_{n,n'}
\\\nn
%&&=1+\ep\, \left[ \left(\frac{\mu(n)\mu(n')}{\gamma(n)\gamma(n')}
%-1\right) -\frac{2\nu\ep}{\sqrt{t}}\,
%\frac{\mu(n)+\mu(n')}{\gamma(n)\gamma(n')} +\frac{\ep^2}{t}\,
%\frac{\nu(2\nu-1)}{\gamma(n)\gamma(n')} +\dots\right]
&&=1+  \left(\frac{\mu(n)\mu(n')}{\gamma(n)\gamma(n')}
-1\right) -\frac{2\nu\ep}{\sqrt{t}}\,
\frac{\mu(n)+\mu(n')}{\gamma(n)\gamma(n')} +\frac{\ep^2}{t}\,
\frac{\nu(2\nu-1)}{\gamma(n)\gamma(n')}
\\\nn&&\equiv 1+ A+B+C \, ,
\eea
\be\label{4.LAt}
\tilde{\Lambda}_{l,l'}	\equiv   \sqrt{\ep}\,  \tilde{\lambda}_{n,n'}
= \sqrt{\ep}\, \frac{2\sqrt{\tau\,t}\sqrt{1-\tau^2}\mu (1+\ep)}
{\gamma(n)\gamma(n')}\,,
\ee
with the notations
\be\label{4.muga}
\mu(n)=\tau\sqrt{t}+2n\sqrt{\ep},\qquad
\gamma(n)=\sqrt{\mu(n)(\mu(n)+\ep/\sqrt{t})}\,.
\ee
Here $\mu(n)$ follows from $l$ and $\ga(n)$ from $l+1$. We divided
$ \Lambda_{l,l'}^{l''}$ into three parts which will be treated
separately in the subsections below. All these will deliver a
contribution of order $\ep$ (including $\ep\ln\ep$ and
$\ep(\ln\ep)^2$) although this  cannot be seen directly from Eq.
\Ref{4.LA}. The same holds for $\tilde{\Lambda}_{l,l'}$, Eq.
\Ref{4.LAt}, after taking the trace over the polarizations.

The  energy is  given by Eq. \Ref{2.E5} and its asymptotic
expansion by Eq.\Ref{3.E1} with another $\cal Z$ which is, of
course, still similar to \Ref{3.Z}. It is expressed in terms of
the matrix $\mathbb{N}_{l,l'}$, Eq.\Ref{2.NED}, by
\be\label{4.Z} {\cal Z}^{\rm }={\rm tr}
    \prod_{i=0}^s   \left( \sqrt{\frac{4\pi t}{\ep}}\,
    \mathbb{N}^{\rm }_{l+\tilde{l}_i,l+\tilde{l}_{i+1}}\right)e^{\eta_{\rm as}}\,.
 \ee
For a matrix $\mathbb{N}_{l,l'}$ we get the asymptotic expansion
\be\label{4.Nlls}
\mathbb{N}_{l,l'}=
\sqrt{\frac{\ep}{4\pi t}}	\ e^{-\eta_{\rm as}}
     \left\{	\left(\begin{array}{cc}1&0\\0&1\end{array}\right)
+
\sqrt{\ep}\,    \left(\begin{array}{cc}
		a^{(\frac{1}{2})(\rm TE)}_{n,n'}&\tilde{\lambda}_{n,n'}\\
		\tilde{\lambda}_{n,n'}&a^{(\frac{1}{2})(\rm TM)}_{n,n'}\end{array}\right)
%\right.  \\ && \left.  ~~~~~
+\ep\, \left(\begin{array}{cc}
		a^{(1)(\rm TE)}_{n,n'}+{\hat{\lambda}}_{n,n'}&0\\
		0&a^{(1)(\rm
TM)}_{n,n'}+{\hat{\lambda}}_{n,n'}\end{array}\right) +\dots
\right\}. \ee
Here the functions $a_{n,n'}$ are the same as in the scalar cases
with the corresponding  boundary conditions.
The factors in front of the figure bracket appear in the
same way as in the scalar case from the Bessel functions.
We remind that in the
given order of the asymptotic expansion the difference between
scalar Neumann boundary conditions and those of the TM mode does
not show up.

In deriving Eq. \Ref{4.Nlls} we have to pay attention
to the $\nu$-dependence of $\lambda_{n,n'}$. Therefore
we defined $\hat{\lambda}_{n,n'}$ by
\be\label{4.lah}
\hat{\lambda}_{n,n'}=
\sqrt{\frac{2\tau}{1+\tau}}
\int\limits_{-\infty}^{\infty}
\frac{d\eta}{\sqrt{\pi}}\, e^{-\eta^2+2i\eta\sqrt{2}\mu+\mu^2}	
%\\&&\nn\times
\sum\limits_{\nu=0}^{\nu_m}
  \frac{\eta^{2\nu}}{\nu!}\, \left(\frac{1-\tau}{1+\tau}\right)^{\nu}
  \, \lambda_{n,n'}.
\ee
This definition is taken in a way that $\lambda_{n,n'}\to 1$
in \Ref{4.lah} gives $\hat{\lambda}_{n,n'} =1$.

Next we have to insert \Ref{4.Nlls} into \Ref{4.Z}. Making a
re-expansion in $\ep$ and taking care of the matrix multiplication
we get
\be\label{4.Z1} {\cal Z}^{\rm as}={\rm tr} \left\{
\left(\begin{array}{cc}1&0\\0&1\end{array}\right) + \ep\left[
\left(\begin{array}{cc}a^{\rm D}&0\\0&a^{\rm N}\end{array}\right)
+ \left(\begin{array}{cc}1&0\\0&1\end{array}\right) \left(
\sum_{i=0}^{s}\hat{\lambda}_{n_i,n_{i+1}}
+\sum_{0<i<j<s}\tilde{\lambda}_{n_i,n_{i+1}}\tilde{\lambda}_{n_j,n_{j+1}}
\right)
 \right] +\dots
\right\}
\ee
In this formula we dropped off-diagonal contributions and those
proportional to $\sqrt{\ep}$ which will disappear later when
carrying out the remaining integrations   as already mentioned in
the preceding section.  In this formula,  the factors resulting
from the T-matrix collected in the same way as in the scalar case
into the functions $a^{\rm D}$ and $a^{\rm N}$. The contributions
from the vector structures \Ref{4.LA} and \Ref{4.LAt} do not
depend on the polarization (see the structure of the translation
formulas, Eq. \Ref{2.trans2}). Therefore these enter proportional
to a unit matrix. Now we take the remaining trace and come to
\be\label{4.Z2} {\cal Z}^{\rm as}= 2 + \left( a^{\rm D}+ a^{\rm N}
+2 \sum_{i=0}^{s}\hat{\lambda}_{n_i,n_{i+1}}
+2\sum_{0<i<j<s}\tilde{\lambda}_{n_i,n_{i+1}}\tilde{\lambda}_{n_j,n_{j+1}}
 \right)\, \ep +\dots
\ee
This expression for $Z^{\rm as}$ must be inserted for $\cal Z$
into the energy, Eq.\Ref{3.E1}. At this place we can already read
off some of the features. First of all, the factor of 2 accounts
for the polarizations of the electromagnetic field. Since the
remaining integrations are the same as for the scalar field we
immediately get the expected result that in PFA the
electromagnetic field gives twice the contribution of a scalar
field. This is the leading order and it is of course independent
of the boundary conditions.

In order $\ep$, the first two factors, $a^{\rm D}$ and $a^{\rm N}$,
 give the same contributions as
in the scalar cases. So we have, for the electromagnetic field,  in the first correction
beyond PFA a contribution with Dirichlet boundary conditions on
both, the sphere and the plane, and another one with Neumann
boundary conditions instead. The corresponding contributions to
the relative corrections beyond PFA to the energy are given by the
first two lines in Eq. \Ref{3.E4} and must be divided by 2.

We mention the role of the minus sign in the latter case. For the
scalar field there were two of them, one following from the
derivative of the Bessel function in the denominator in
\Ref{2.dN}. This sign has a corresponding one in the
electromagnetic case in the denominator of \Ref{2.dTM}. The other
minus sign followed in the scalar case from the sign in the
logarithm in Eq.\Ref{2.E3}. In the electromagnetic case the
corresponding one is the minus sign in front of $d^{\rm TM}_l$ in
\Ref{2.NED}.

The remaining   terms in Eq.\Ref{4.Z2} just represent the
additional contributions which come in from the vector character
of the electromagnetic field, i.e., from its spin. There are
contributions diagonal in the polarizations, resulting from
$\Lambda_{l,l'}^{l''}$, Eq.\Ref{2.LA}, and off-diagonal ones resulting from
$\tilde{\Lambda}_{l,l'}$, Eq.\Ref{2.LAt}. It should be mentioned
that in the considered first order in $\ep$ all these
contributions enter additivly. A mixing of these will happen in  higher orders
only.

In the following subsections we consider separately the
contributions resulting from the three parts, A, B and C of
$\Lambda_{l,l'}^{l''}$, Eq.\Ref{4.LA} and that of
$\tilde{\Lambda}_{l,l'}$, Eq.\Ref{4.LAt}. Starting from part B we
will meet expressions where it is not possible to make a direct
expansion for small $\ep$. For illustration we consider a simple
example. Consider the integral
\be\label{4.example1} f(\ep)=\int\limits_0^1 d\tau\,\frac{
g(\tau)}{\tau+a \ep+b\ep^2} \ee
for $\ep\to0$. In case the function $g(\tau)$ has a zero,
$g(0)=0$, we can put $\ep=0$ directly under the sign of the
integral. In opposite, if $g(0)\ne0$ holds, we cannot do that
since the $\tau$-integration would diverge. The only we can do is
to put $\ep=0$ where it goes with the coefficient $b$,
\be\label{4.example2} f(\ep)=\int\limits_0^1 d\tau\,\frac{
g(\tau)}{\tau+a \ep } +\dots\,, \ee
where the dots denote contributions of higher order in $\ep$. The
remaining integral must be treated in some other way. For example, we can
 integrate by parts and   expand after that,
\bea\label{4.example3} f(\ep)&=&\ln(1+\ep)\,g(1)-\ln\ep\,
g(0)-\int_0^1d\tau\ln(\tau+\ep) g'(\tau)+\dots\,, \nn\\&=& -\ln\ep
\,g(0)-\int_0^1d\tau\, \ln\tau \ g'(\tau)+\dots\,. \eea
Below such and similar situations will appear repeatedly.

%%%%%%%%%%%%%%%%%%%%%%%%%%%%%%%%%%%%%%%%%%%%%%%%%%%%%%%%%%%%%%%%%%%%%%%%%%%
%%%%%%%%%%%%%%%%%%%%%%%%%%%%%%%%%%%%%%%%%%%%%%%%%%%%%%%%%%%%%%%%%%%%%%%%%%%
%%%%%%%%%%%%%%%%%%%%%%%%%%%%%%%%%%%%%%%%%%%%%%%%%%%%%%%%%%%%%%%%%%%%%%%%%%%
\subsection{Part A in the $\Lambda$-contribution}
Part A is given by
\be\label{4.1.A1}
A=\frac{\mu(n)\mu(n')}{\gamma(n)\gamma(n')}-1
\ee
with $\mu(n)$ and $\gamma(n)$ defined in Eq. \Ref{4.muga}.
The contribution from this part to the energy
is quite easy to calculate since it is possible to expand $A$
 directly in powers of $\ep$,
\be\label{4.1.A2} A=-\frac{1}{t\tau}\,\ep+\dots\,,
\ee
without causing any divergences.
We define its contribution to $\lambda_{n,n'}$ by
\be\label{4.1.la}
\lambda_{A;n,n'}=-\frac{1}{t\tau}.
\ee
This is a quite simple formula, for instance the dependence on $n$
and $n'$  dropped out. Further, since it does not depend on $\nu$,
from \Ref{4.lah} we have $\hat{\lambda}_{A;n,n'}=\lambda_{A;n,n'}$
and   its contribution to
   ${\cal Z}^{\rm as}$, \Ref{4.Z2},  is
\be\label{4.1.1}
{\cal Z}^{\rm as}_{\rm }=2-2\ep\,\sum_{i=0}^s\frac{1}{t\tau}+\dots\,.
\ee
Here the dots stand for all  other contributions.  We denote the
corresponding part of the energy by $\Delta E_A$ and we
come with \Ref{3.E1} to
\bea\label{4.1.E1}
E_{\rm PFA}+\Delta E_{\rm A}&=&
\frac{R}{4\pi d^2}
\sum_{s=0}^\infty \frac{-2}{s+1}
\int_0^\infty dt \, t  \
\int_{0}^1 \frac{d\tau\,\sqrt{\tau}}{\sqrt{1-\tau^2}}\,
\int_{-\infty}^\infty \frac{d\mu}{\sqrt{\pi}}\,
\nn \\ &&
\times \left(\prod_{j=1}^s\int_{-\infty}^\infty
\frac{dn_j}{\sqrt{\pi}}\right) \
 \left(1-\ep\,\frac{s+1}{\tau\, t}+\dots
\right) e^{-2t(s+1)- \mu^2(s+1)/{\tau}   -\eta_1}.
\eea
We mention that we have taken the lower integration limit in the
$n_j$-integrations to $-\infty$ which is possible since the
remaining integrations do converge. The integrations in this
expression can be carried out easily and the result is
\bea\label{4.1.EA} E_{\rm PFA}+\Delta E_{\rm A}&=& \frac{-2R}{4\pi
d^2 }\left(\frac{\zeta(4)}{4}-\ep\,\frac{\pi
\zeta(2)}{4}+\dots\right)\,. \eea
The correction resulting from part A is
\be\label{4.1.EA1}
\Delta E_{\rm A}=\frac{R}{4\pi d^2}\ \frac{\pi \zeta(2)}{2}\ \ep\,.
%\Delta E_{\rm A}}{E_{\rm PFA}}=-\frac{15}{\pi}\,\ep
 \ee
%

%
%
%\nn\\
%&=&E_{\rm PFA}\left(1-\ep\,\frac{15}{\pi}+\dots\right). \eea
%Finally we define   the relative correction
%%
%\be\label{4.1.eA}\Delta e_{\rm A}=\frac{\Delta E_{\rm A}}{E_{\rm PFA}}=-\frac{15}{\pi}\,\ep
% \ee
%%
%resulting from part A.
%%%%%%%%%%%%%%%%%%%%%%%%%%%%%%%%%%%%%%%%%%%%%%%%%%%%%%%%%%%%%%%%%%%%%%%%%%%
%%%%%%%%%%%%%%%%%%%%%%%%%%%%%%%%%%%%%%%%%%%%%%%%%%%%%%%%%%%%%%%%%%%%%%%%%%%
%%%%%%%%%%%%%%%%%%%%%%%%%%%%%%%%%%%%%%%%%%%%%%%%%%%%%%%%%%%%%%%%%%%%%%%%%%%
\subsection{Part B in the $\Lambda$-contribution}
Part B is given by
\be\label{4.2.B1}
B=\frac{-2\nu\ep}{\sqrt{t}}\frac{\mu(n)+\mu(n')}{\gamma(n)\gamma(n')}.
\ee
Regrettably, its contribution to the energy cannot be calculated
so easy as before. First of all we have an additional dependence
on $\nu$. In addition, it is impossible to make a
simple expansion in $\ep$. This would produce a singularity in the
$\tau$-integration.

We define $\lambda_{B;n,n'}=B/\ep$ which must be inserted into \Ref{4.lah}.
The summation over $\nu$ is quite simple,
\be\label{4.2.nu1}
\sum\limits_{\nu=0}^{\infty}\,\nu\,
  \frac{\eta^{2\nu}}{\nu!}\, \left(\frac{1-\tau}{1+\tau}\right)^{\nu}
  =\frac{1-\tau}{1+\tau}\,\eta^2\,\exp\left(\frac{1-\tau}{1+\tau}\,\eta^2\right).
\ee
Here we have taken $\nu_m=\infty$ since this does not cause
singularities. Next we need to carry out the
integration over $\eta$ in \Ref{3.Nlls}. It is Gaussian,
\be\label{4.2.eta}
\int\limits_{-\infty}^{\infty} \frac{d\eta}{\sqrt{\pi}}\,
\frac{1-\tau}{1+\tau}\,\eta^2\,
\exp\left({\frac{1-\tau}{1+\tau}\,\eta^2-\eta^2+2i\eta\sqrt{2}\mu+\mu^2}\right)
=h_B(\tau,\mu)
\sqrt{\frac{1+\tau}{2\tau}}
\,
\exp\left(-\frac{\mu^2}{\tau}\right).
\ee
with
\be\label{4.2.h}h_B(\tau,\mu)=
\frac{1-\tau}{4\tau}\left(1-2\frac{1+\tau}{\tau}\,\mu^2\right).
\ee
Using Eq.\Ref{4.2.eta} in \Ref{4.lah} we get
\be\label{4.2.lah}
\hat{\lambda}_{B;n,n'}=h_B(\tau,\mu)\,\lambda_{B;n,n'}.
\ee
%
%In this way, the corresponding
%contribution to the function $N^{\rm as}_{l,l'}$, Eq.\Ref{3.Nlls}, which we
%denote by $N^{\rm as}_{{\rm B};l,l'}$, is
%%
%\be\label{4.2.NB}
%N^{\rm as}_{{\rm B};l,l'}=
%\frac{-2}{\sqrt{t}}  \sqrt{\frac{\ep}{4\pi t}}
%\, h(\tau,\mu)\,
%      \frac{\mu(n)+\mu(n')}{\gamma(n)\gamma(n')}
%      \,e^{-2t-\mu^2/\tau-(n-n')^2}.
%\ee
%%
%%
%It must be inserted into $\cal Z$, Eq.\Ref{4.Z}. The prefactors
%cancel as before and the result is the contribution from part B to
%$\tilde{\lambda}_{n,n'}$,
%%
%\be\label{4.2.lat}
%\tilde{\lambda}_{n,n'}=\frac{-2}{\sqrt{t}}\,h(\tau,\mu) \,\frac{\mu(n)+\mu(n')}{\gamma(n)\gamma(n')}.
%\ee
%%
This must be inserted into $\cal Z^{\rm as}$, \Ref{4.Z2}, and further into the energy.
With \Ref{3.E1}   it is
\bea\label{4.2.E1}
\Delta E_{\rm B}&=&
-\frac{R}{4\pi d^2}\sum_{s=0}^\infty\frac{-4\ep}{s+1}
\int_0^\infty dt\,\sqrt{t}\,e^{-2t(s+1)}
  \,\tilde{B}\,,
\eea
where we defined
\be\label{4.2.Bt1}
\tilde{B}=
 \int_0^1\frac{d\tau\,\sqrt{\tau}}{\sqrt{1-\tau^2}}
 \int_{-\infty}^\infty \frac{d\mu}{\sqrt{\pi}}\ e^{-\mu^2(s+1)/\tau}\
 h_B(\tau,\mu)\, \left(\prod_{j=1}^s \int_{n_0}^\infty\frac{dn_j}{\sqrt{\pi}} \right)
  \sum_{i=0}^s \frac{\mu(n_i)+\mu(n_{i+1})}{\gamma(n_i)\gamma(n_{i+1})}
  \,e^{-\eta_1}\,.
\ee
The integration over $\mu$ is Gaussian and we get
\bea\label{4.2.Bt2} \tilde{B}=\frac{1}{\sqrt{s+1}}
 \int_0^1d\tau\,f_B(\tau)
\left(\prod_{j=1}^s \int_{n_0}^\infty\frac{dn_j}{\sqrt{\pi}} \right)
  \sum_{i=0}^s \frac{\mu(n_i)+\mu(n_{i+1})}{\gamma(n_i)\gamma(n_{i+1})}
   \,e^{-\eta_1}\,
\eea
with
\be\label{4.2.f}
f_B(\tau)=\frac{1}{4}\sqrt{\frac{1-\tau}{1+\tau}}\left(1-\frac{1+\tau}{s+1}\right).
\ee
%
%The function $\tilde{B}$ can be simplified using the symmetry under $i\to s-i$
%in the sum over $i$,
%%
%\bea\label{4.2.Bt3} \tilde{B}=
%\frac{2}{\sqrt{s+1}}\int_0^1d\tau\,f_B(\tau) \left(\prod_{j=1}^s
%\int_{n_0}^\infty\frac{dn_j}{\sqrt{\pi}} \right)
%  \sum_{i=0}^s \frac{\mu(n_i)}{\gamma(n_i)\gamma(n_{i+1})}\,.
%\eea
%%
Now we have still $\ep$ in a number of places in \Ref{4.2.Bt2}. Still we
cannot simply expand for small $\ep$. If we would do so, because
of
\be\label{4.2.mu1}
 \frac{\mu(n_i)}{\gamma(n_i)\gamma(n_{i+1})}=\frac{1}{\tau\sqrt{t}}+\dots\,,
\ee
where we used \Ref{4.muga},   the $\tau$-integration would become
logarithmically  divergent. The only places where we can put
$\ep=0$ directly are those where it goes with $n$ in
$\mu(n)$ and in $\gamma(n)$,
\be\label{4.2.mu2} \frac{\mu(n_i)}{\gamma(n_i)\gamma(n_{i+1})}
%=\sqrt{\frac{\tau\sqrt{t}+2n_i\sqrt{\ep}}{\tau\sqrt{t}+2n_i\sqrt{\ep}+\ep\sqrt{t}}}
%\frac{1}{\sqrt{\tau\sqrt{t}(\tau\sqrt{t}+2n_{i+1}\sqrt{\ep}+\ep\sqrt{t})}}
=\frac{1}{\sqrt{\tau\sqrt{t}(\tau\sqrt{t}+\ep/\sqrt{t})}}+\dots
\,. \ee
Also we can take $n_0=-\infty$. In doing so we do not produce divergences in the integrations.
After that we are left with a simpler $\tau$-integration. Here we
integrate by parts,
\be\label{4.2.dtau} \int_0^1d\tau\,
\frac{f_B(\tau)}{\sqrt{\tau\sqrt{t}(\tau\sqrt{t}+\ep/\sqrt{t})}} =
-\frac{2}{\sqrt{t}}\int_0^1d\tau\,
\ln\frac{\sqrt{\tau\sqrt{t}}+\sqrt{\tau\sqrt{t}+\ep/\sqrt{t}}}
{\sqrt{\ep / \sqrt{t}}} \frac{\pa}{\pa \tau}\, f_B(\tau)\,. \ee
The surface term is zero. In the new $\tau$-integral it is
possible to expand the logarithm for small $\ep$. We insert the
result into $\tilde{B}$, Eq.\Ref{4.2.Bt2}, and get
\be\label{4.2.dtau1} \tilde{B} = \frac{-4}{\sqrt{t}} \left[
\left(-\frac{1}{2}\ln\ep+\frac{1}{2}\ln t+\ln
2\right)\left(f_B(1)-f_B(0)\right)+\frac{1}{2} \int_0^1d\tau\,
\ln\tau \frac{\pa}{\pa \tau}\, f_B(\tau) \right]+\dots\,. \ee
%
%%
%\be\label{4.2.dtau} \int_0^1d\tau\,
%\frac{f(\tau)}{\sqrt{\tau\sqrt{t}(\tau\sqrt{t}+\ep\sqrt{t})}}
%=
%-\frac{2}{\sqrt{t}}
%\left[
%\left(-\frac{1}{2}\ln\ep+\frac{1}{2}\ln t+\ln 2\right)\left(f(1)-f(0)\right)+\frac{1}{2}
%\int_0^1d\tau\,
%\ln\tau
%\frac{\pa}{\pa \tau}\, f(\tau)
%\right]\,.
%\ee
%%
Since after \Ref{4.2.mu2} there is no more any $n_j$-dependence
 we used formula \Ref{66} and accounted
also for the sum over $i$. The remaining integration over $\tau$ can  be
carried out easily,
\be\label{4.2.f1} \int_0^1d\tau\, \ln\tau \frac{\pa}{\pa \tau}\,
f_B(\tau) =
%\frac{(1+s)\pi-2-s\ln4}{8(1+s)}\,.
\frac{\pi}{8}-\frac{1+s\ln2}{4(s+1)}\,.
\ee
We mention that Eq.\Ref{4.2.dtau1} is  the first place where a logarithm
in $\ep$ appears. Its origin is
clearly seen from Eq.\Ref{4.2.mu1}.

Next we have to insert $\tilde{B}$ into the energy \Ref{4.2.E1}.
The remaining $t$-integration is now a bit more complicated since
it involves  a $\ln t$. From that a logarithm $\ln(1+s)$ appears.
However, simple calculations yield
\be\label{4.2.EB} \Delta E_{\rm  B}=
\frac{R}{4\pi d^2}\left[
\left(\zeta(3)-\zeta(2)\right)(\gamma-2\ln2)
+\zeta'(2)-\zeta'(3)
  -\frac{\pi}{2}\,\zeta(2)+\zeta(3)
+\frac{1}{4\pi}\left(\zeta(3)-\zeta(2)\right)\,\ln\ep\
\right]\,\ep\,.
\ee
This is the contribution from part B in \Ref{} to the corrections beyond PFA.
It involves a logarithm in $\ep$ and it has an analytic expression.
%%%%%%%%%%%%%%%%%%%%%%%%%%%%%%%%%%%%%%%%%%%%%%%%%%%%%%%%%%%%%%%%%%%%%%%%%%%
%%%%%%%%%%%%%%%%%%%%%%%%%%%%%%%%%%%%%%%%%%%%%%%%%%%%%%%%%%%%%%%%%%%%%%%%%%%
%%%%%%%%%%%%%%%%%%%%%%%%%%%%%%%%%%%%%%%%%%%%%%%%%%%%%%%%%%%%%%%%%%%%%%%%%%%
\subsection{Part C in the $\Lambda$-contribution}
Part C is given by
\be\label{4.3.C}
C=\frac{\ep^2}{t}\frac{\nu(2\nu-1)}{\gamma(n)\gamma(n')}. \ee
The calculation of its contribution to the energy requires most
effort. First of all we   observe a quadratic dependence on
$\nu$. We have to insert \Ref{4.3.C} into \Ref{3.Nlls} and define
the corresponding contribution to $N_{l,l'}^{\rm as}$ by
\bea\label{4.3.NC} N_{C;l,l'}^{\rm as}= &=&\sqrt{\frac{\ep
\tau}{2\pi t(1+\tau)}}\
e^{-2t-(n-n')^2}\int\limits_{-\infty}^{\infty}
\frac{d\eta}{\sqrt{\pi}}\, e^{-\eta^2+2i\eta\sqrt{2}\mu+\mu^2}	
\\&&\nn\times
\sum\limits_{\nu=0}^{\nu_m}
  \frac{\eta^{2\nu}}{\nu!}\, \left(\frac{1-\tau}{1+\tau}\right)^{\nu}
C.
\eea
Here it is impossible to take the upper limit $\nu_m$ of the
summation over $\nu$ to infinity. Doing so would produce a factor
$\tau^{-2}$ and making the $\tau$-integration diverge. Therefore
we must account for a finite $\nu_m$, Eq.\Ref{3.num}. With the
substitution \Ref{3.subst} it becomes
\be\label{4.3.num} \nu_m=\frac{t\tau}{\ep}+\dots\,,
\ee
Technically we account for it by inserting a step function into
the sum and formally summing up to infinity as before. For the
step function we take the integral representation
\be\label{4.3.step}
\Theta(\nu_m-\nu)=\int_{-\infty}^\infty\frac{dp}{2\pi i}\,
\frac{\exp\left(ip\left(\frac{t\tau}{\ep}-\nu\right)\right)}{p-i0}
\ee
and define
\be\label{4.3.Nt}
{N}_{C;l,l'}^{\rm as}=
\int_{-\infty}^\infty\frac{dp}{2\pi i}\,
\frac{\exp\left(ip \frac{t\tau}{\ep} \right)}{p-i0}\,
\tilde{N}_{C;l,l'}^{\rm as}
 \ee
The summation over $\nu$ now involves the additional factor $\al^\nu$ with
\be\label{4.3.al}\al\equiv e^{-i p}. \ee
This sum and also the integration over $\eta$ can be done generalizing \Ref{4.2.nu1} and \Ref{4.2.eta}.
The result is
\be\label{4.3.Nt1}
\tilde{N}_{C;l,l'}^{\rm as}=
\frac{\ep^2}{t}\frac{\LA^{1/2}}{\gamma(n)\gamma(n')}\,
h_C(\tau,\mu) \, \sqrt{\frac{\ep}{4\pi t}}\, e^{-\eta_{\rm as}-\mu^2/(\tau\LAt)}
\ee
with
\bea\label{4.3.hC}
h_C(\tau,\mu)&=&
6\left(\frac{1-\tau}{4\tau}\,\al\,\LA\right)^2
\left[1-4\frac{1+\tau}{\tau}\,\LA\,\mu^2
+\frac{4}{3}\left(\frac{1+\tau}{\tau}\,\LA\,\mu^2\right)^2\right]
\nn\\&&
+\frac{1-\tau}{4\tau}\,\al\,\LA
\left[1-2\frac{1+\tau}{\tau}\,\LA\,\mu^2
\right].
\eea
Here we defined (for use only in this subsection)
\be\label{4.3.LA}
\LA=\frac{2\tau}{1+\tau-\al(1-\tau)},\qquad \LAt=\frac{2}{2-(1-\tau)(1-\al)}.
\ee
From  \Ref{4.3.Nt} we can now read off the contribution from part
C into $\tilde{\lambda}_{n,n'}$ and insert that into $\cal Z^{\rm
as}$, Eq.\Ref{4.Z1} and further into the energy \Ref{3.E1}. The
corresponding contribution is
\bea\label{4.3.E1} \Delta E_{\rm C}&=& -\frac{R}{4\pi  d^2}
\sum_{s=0}^\infty\frac{2\ep^2}{s+1}
\int_0^\infty dt\, e^{-2t(s+1)}
  \,\tilde{C}\,,
\eea
where we defined
\be\label{4.3.Ct}     \tilde{C}=
 \int_0^1\frac{d\tau\,\sqrt{\tau}}{\sqrt{1-\tau^2}}
 \int_{-\infty}^\infty \frac{d\mu}{\sqrt{\pi}}\ e^{-\mu^2(s+1)/\tau}
\left(\prod_{j=1}^s \int_{n_0}^\infty\frac{dn_j}{\sqrt{\pi}} \right)
  \sum_{i=0}^s \frac{e^{-\eta_1}}{\gamma(n_i)\gamma(n_{i+1})}\,
\int_{-\infty}^\infty\frac{dp}{2\pi i}\,
\frac{\exp\left(ip \frac{t\tau}{\ep} \right)}{p-i0}\,h_C(\tau,\mu)\,.
\ee
Here, again, the integration over $\mu$ is Gaussian and we come to
\be\label{4.3.Ct1}     \tilde{C}=\frac{1}{\sqrt{s+1}}
 \int_0^1\frac{d\tau\,{\tau}}{\sqrt{1-\tau^2}}
\left(\prod_{j=1}^s \int_{n_0}^\infty\frac{dn_j}{\sqrt{\pi}} \right)
  \sum_{i=0}^s \frac{e^{-\eta_1}}{\gamma(n_i)\gamma(n_{i+1})}\,
R(\tau,t)\ee
with
\be\label{4.3.R} R(\tau,t)=
\int_{-\infty}^\infty\frac{dp}{2\pi i}\,
\frac{\exp\left(ip \frac{t\tau}{\ep} \right)}{p-i0}\, \left(\LA
\LAt\right)^{1/2}\left[6\left(\al \LA f_C(\tau)\right)^2+ \al \LA
f_C(\tau)\right] \ee
and
\be\label{4.3.fC}
f_C(\tau)=\frac{1-\tau}{4\tau}\left(1-\frac{1+\tau}{1+s}\LA\LAt\right).
\ee
In Eq.\Ref{4.3.Ct1}, the main contribution comes from $\tau\sim
0$. This is because the function $f_C(\tau)$ diverges like
$f_C(\tau)\sim \tau^{-2}$ for $\tau\to0$. The integration over
$\tau$ is nevertheless finite  for   $\ep\ne 0$ because of the
exponential involving $p$. Next we want to carry out the
integration over $p$ in \Ref{4.3.R}. For this we move the
integration contour upwards in the complex $p$-plane. We have a
pole at $p=0$. Its contribution is easy to calculate and it gives
just the result we would obtain with putting $\nu_m=\infty$ at the
very beginning. Further there are poles in
$p_0=i\ln\frac{1+\tau}{1-\tau}+2\pi n$ ($n$ integer) from $\LA$
and cuts starting from $p_c=i\ln\frac{1+\tau}{1-\tau}+\pi+2\pi n$
resulting from $\LA\LAt$.

For small $\ep$, non-vanishing contributions come only from the
poles in $p=0$ and in $p=p_0$ with $n=0$. For these we can expand
$p_0=2i\tau+\dots$ and $\LA\LAt=1+\dots$ and $R$ takes the form
\be\label{4.3.R1}
R(\tau,t)=
\int_{-\infty}^\infty\frac{dp}{2\pi i}\,
\frac{\exp\left(ip \frac{t\tau}{\ep} \right)}{p-i0}\,
\left[6\left(\al \frac{-p_0}{p-p_0}\, f_C(\tau)\right)^2
+ \al  \frac{-p_0}{p-p_0}\,f_C(\tau)\right]+\dots\,.
\ee
The pole in $p_0$ is second order and we get
\be\label{4.3.R2}
R(\tau,t)=6f_C(\tau)^2\left[1-\left(1+\frac{2t\tau^2}{\ep}\right)\,e^{-2t\tau^2/\ep}\right]+
f_C(\tau)\left[1- e^{-2t\tau^2/\ep}\right]+\dots\,.
\ee
We mention that the first term in both square brackets result from
a $\nu$-summation with $\nu_m=\infty$. The terms with the
exponentials result from taking a finite $\nu_m$.
As a result, while $\ep\ne0$, the function $R(\tau,t)$
is finite for $\tau\to0$ instead of diverging like $\tau^{-2}$ as
the function $f_C(\tau)$ does. We emphasize that this is the
justification for accounting for a finite $\nu_m$ in part C. In
the other parts, because we did not have a small-$\tau$ behavior
like in this one, we could take $\nu_m=\infty$ without hitting a
divergence.

The expression for $R$, Eq.\Ref{4.3.R2}, must be inserted into
$\tilde{C}$, Eq.\Ref{4.3.Ct1}. To proceed with the expansion for
small $\ep$ we observe that we cannot put $\ep=0$ in $R$ since
this would return us to a situation where the $\tau$-integration
diverges at $\tau\to 0$. Instead we make in $\tilde{C}$ the
substitution $\tau\to\tau \sqrt{\ep}$,
\be\label{4.3.Ct2}
\tilde{C}=
\frac{1}{\ep}\frac{1}{\sqrt{s+1}}
\int_0^{1/\sqrt{\ep}}\frac{d\tau\,\tau}{\sqrt{1-\tau^2\ep}}
\left(\prod_{j=1}^s \int_{\tilde{n}_0}^\infty\frac{dn_j}{\sqrt{\pi}} \right)
  \sum_{i=0}^s \frac{e^{-\eta_1}}{\tilde{\gamma}(n_i)\tilde{\gamma}(n_{i+1})}\,
 R(\tau \sqrt{\ep},t)
\ee
with
\be\label{4.3.ga}
\tilde{\gamma}(n)=\sqrt{\left(\tau\sqrt{t}+2n\right)
\left(\tau\sqrt{t}+2n+\sqrt{\frac{\ep}{t}}\right)}\,
\ee
and
\be\label{4.3.n0t}
\tilde{n}_0=-\frac{1}{2}\tau\sqrt{t}\,.
\ee
In this way, the $\tau$-integration produces a factor $1/\ep$
which makes the contribution from part C, which initially went
with a factor $\ep^2$,  a contribution first order in $\ep$. We
mention that it resulted from the factors $\gamma(n)$ in the
denominator.

There is still a dependence on $\ep$ in $\tilde{C}$,
Eq.\Ref{4.3.Ct2}. It is twofold. First is that which goes with
$\tau$. Here we can put $\ep=0$. The same we can do in the upper
integration limit.
We note
\be\label{4.3.R3}
R(\tau\sqrt{\ep},t)=\frac{3}{2\ep}\left(\frac{s}{s+1}\right)^2g(2t\tau^2)+O(1)
\ee
with
\be\label{4.3.g} g(x)=\frac{1}{4}\left(1-(1+x)\, e^{-x}\right)\,
\ee
for $\ep\to0$ and for $\tilde{C}$ we get
\be\label{4.3.Ct3} \tilde{C}=
\frac{3}{2\ep} \frac{s^2}{(s+1)^{5/2}}
\int_0^{\infty}\frac{d\tau}{\tau}    \, g(2t\tau^2)
\left(\prod_{j=1}^s
\int_{\tilde{n}_0}^\infty\frac{dn_j}{\sqrt{\pi}} \right)
  \sum_{i=0}^s \frac{e^{-\eta_1}}{\tilde{\gamma}(n_i)\tilde{\gamma}(n_{i+1})}\,.
\ee
In order to simplify the representation of $\tilde{C}$ we make in
\Ref{4.3.Ct3} the substitution $\tau\to\tau/\sqrt{t}$. After that
it takes the form
\be\label{4.3.Ct4}
\tilde{C}=\sigma\int_0^\infty\frac{d\tau}{\tau}g(2\tau^2)
\left(\prod_{j=1}^s \int_{\hat{n}_0}^\infty\frac{dn_j}{\sqrt{\pi}}
\right)
  \sum_{i=0}^s
  \frac{e^{-\eta_1}}{\hat{\gamma}(n_i)\hat{\gamma}(n_{i+1})}\,,
\ee
where we introduced the notations
\be\label{4.3.notat}
\sigma=\frac{3}{2\ep}\frac{s^2}{(s+1)^{5/2}}\,,\qquad
\hat{n}_0=-\frac{\tau}{2}\,,\qquad
\hat{\gamma}(n)=\sqrt{(\tau+2n)(\tau+2n+\sqrt{\ep/{t}})}\,, \ee
which will be used in the remaining part of this subsection.

In this way we are left with the dependence on $\ep$ in
$\hat{\gamma}$. Here we can not put directly $\ep=0$. This  would
produce a   divergence in the integrations at $\tau=-2n$ and an
imaginary part would appear which is clearly not present in the
energy. The way out is a partial integration in the $n_i$ and in
the $n_{i+1}$ integrations. But before we can do that we have to
pay attention to the contributions from $i=0$ and from $i=s$ in
the sum over $i$. From the formal setting in Eq.\Ref{3.Z} we have
to put  $n_0=n_s=0$ in Eq.\Ref{4.3.Ct4} and the corresponding
$\hat{\gamma}$ do not depend on any $n$. We denote the
contributions from $i=0$ and $i=s$ (both give for symmetry reasons
the same contribution) by $\tilde{C}_0$ and the remaining one,
i.e., that for $i=1,\dots,s-1$, by $\tilde{C}_1$.

First we consider $\tilde{C}_0$. Since the function $g(x)\sim x^2$
for $x\to0$ we can take $\hat{\gamma}(0)=\tau+\dots$ and are left
with
\be\label{4.3.Ct0.1}
\tilde{C}_0=
2\sigma\int_0^\infty\frac{d\tau}{\tau^2}\,g(2\tau^2)
\left(\prod_{j=1}^s \int_{\hat{n}_0}^\infty\frac{dn_j}{\sqrt{\pi}}
\right)
  \frac{e^{-\eta_1}}{\hat{\gamma}(n_1)}\,.
\ee
Now we integrate by parts according to
%%
%\bea\label{4.3.pi}
%\int_{-\tau/2}^\infty dn\, \frac{e^{-\eta_1}}{\hat{\gamma}(n)}
%&=&
%\ln\left(2\left(\sqrt{\tau+2n}+\sqrt{\tau+2n+\sqrt{\ep/t}})\right)\right)
%\,e^{-\eta_1}\bigg|_{-\tau/2}^{\infty}
%\nn\\&&
%-\int_{-\tau/2}^\infty dn\,
%\ln\left(2\left(\sqrt{\tau+2n}+\sqrt{\tau+2n+\sqrt{\ep/t}})\right)\right)\frac{\pa}{\pa n}\,e^{-\eta_1}
%\eea
%%
%
\bea\label{4.3.pi}
\int_{-\tau/2}^\infty dn\, \frac{e^{-\eta_1}}{\hat{\gamma}(n)}
&=&
-\ln\left(2\left(\frac{\ep}{t}\right)^{1/4}\right)
\,e^{-\eta_2}
\nn\\&&
-\int_{-\tau/2}^\infty dn\,
\ln\left(2\left(\sqrt{\tau+2n}+\sqrt{\tau+2n+\sqrt{\ep/t}})\right)\right)\frac{\pa}{\pa n}\,e^{-\eta_1}
\eea
with
\be\label{4.3.pi1}\eta_2={\eta_1}_{|n=-\tau/2}. \ee
Here we can expand for small $\ep$, \bea\label{4.3.pi2}
\int_{-\tau/2}^\infty dn\, \frac{e^{-\eta_1}}{\hat{\gamma}(n)} &=&
-\ln\left(2\left(\frac{\ep}{t}\right)^{1/4}\right) \,e^{-\eta_2}
-\int_{-\tau/2}^\infty dn\,
\ln\left(\sqrt{\tau+2n}\right)\frac{\pa}{\pa
n}\,e^{-\eta_1}+\dots\,. \eea
This must be inserted into Eq.\Ref{4.3.Ct0.1}. After that we make
there the substitution $n_j\to n_j\, \tau/2$. This allows to
change the orders of integrations,
\bea\label{4.3.Ct0.2}
\tilde{C}_0
&=&
\frac{-2\sigma}{\sqrt{\pi}}
 \left(\frac{1}{4}\ln\frac{\ep}{t}+\ln2\right)
R(s)
%\nn \\ &&
+{4\sigma}
Q(s)\,,
\eea
and we introduced the notations
\bea\label{4.3.Rs}
R(s)&=&\left(\prod_{j=2}^s \int_{-1}^\infty\frac{dn_j}{\sqrt{\pi}}\right)
     g_1\left(\frac{\eta_2}{4}\right)
\nn \\
Q(s)&=&
\left(\prod_{j=1}^s \int_{-1}^\infty\frac{dn_j}{\sqrt{\pi}}\right)\,(2n_1-n_2)\,
      g_2\left(\frac{\eta_1}{4}\right)
\eea
with
\bea\label{4.3.g1}
g_1(x)&=&\int_0^\infty\frac{d\tau}{\tau^2}\,g(2\tau^2)\left(\frac{\tau}{2}\right)^{s-1}\, e^{-x \tau^2}
,\nn\\
g_2(x)&=&\int_0^\infty\frac{d\tau}{\tau^2}\,g(2\tau^2)
\ln\left(4\sqrt{\tau}\sqrt{1+2n_1}\right)
\left(\frac{\tau}{2}\right)^{s+1}\, e^{-x \tau^2}.
\eea
The  integrations over $\tau$ can be carried out explicitly, however the
formulas are too voluminous as to be displayed here. The remaining
integrations over the $n_j$ can be performed only numerically.
Even that is not an easy task since the integrals are $s$-dimensional.
 We could proceed only till $s=5$ for $R(s)$ and $s=3$ for $Q(s)$. However, since the sum over
 $s$ is quite fast converging this gives at least the order of magnitude correctly.

Expression \Ref{4.3.Ct0.2} must be inserted into the energy
\Ref{4.3.E1}. Introducing the corresponding notation it is
\be\label{4.3.Ect0} \Delta E_{\tilde{C}_0} =-\frac{R}{4\pi  d^2}
\sum_{s=0}^\infty\frac{2\ep^2}{s+1} \int_0^\infty dt\,
e^{-2t(s+1)} \tilde{C}_0\,. \ee
Here, the $t$-integration can be carried out easily and summing
over $s$, as far as data are available, the result is
\be\label{4.3.Ect0_1}
\Delta E_{\tilde{C}_0}=\frac{R}{d^2}\, \ep\left( 0.0020 + 0.00017 \,\ln\ep\right).
\ee

Now we have to consider $\tilde{C}_1$, i.e., the contributions from $i=1,\dots,s-1$ to \Ref{4.3.Ct4}.
We make the substitution $n_j\to n_j\,\tau/2$ such that it takes the form
\be\label{4.3.Ct1.1} \tilde{C}_1=
\sigma\int_0^\infty\frac{d\tau}{\tau^2}\,g(2\tau^2)\left(\frac{\tau}{2}\right)^{s}
\left(\prod_{j=1}^s \int_{-1}^\infty\frac{dn_j}{\sqrt{\pi}}
\right)  \sum_{i=1}^{s-1}
  \frac{e^{-\eta_1 \tau^2/4}}{\bar{\gamma}(n_i)\bar{\gamma}(n_{i+1})}\,
\ee
with $\bar{\gamma}(n)=\sqrt{(1+n)(\tau(1+n)+\sqrt{\ep/t}}$.
For the  integration by parts we adopt the following scheme,
\bea\label{4.3.pi3}
\int_{-1}^\infty dn\, \frac{e^{-\eta_1\tau^2/4}}{\bar{\gamma}(n)}
&=&
-\frac{1}{\sqrt{\tau}}\left[
\frac14\ln\frac{\ep}{\tau}-\ln \left(2\sqrt{\tau}\right)\right]
\,e^{-\eta_2\tau^2/4}
\nn\\&&
-\frac{1}{\sqrt{\tau}}\int_{-1}^\infty dn\,
\ln\frac{\sqrt{\tau}\sqrt{1+n}+\sqrt{\tau(1+n)+\sqrt{\ep/t}}}{\sqrt{\tau}+\sqrt{\tau+\sqrt{\ep/t}}}
\frac{\pa}{\pa n}\,e^{-\eta_1\tau^2/4}\,.
\eea%\nn\\&=&
Here we can take $\ep\to0$ and get
\bea\label{4.3.pi4}
\int_{-1}^\infty dn\, \frac{e^{-\eta_1\tau^2/4}}{\hat{\gamma}(n)}
=
-\frac{1}{\sqrt{\tau}}
\int_{-1}^\infty dn\,
\left( L_1(n)+L_2(n)\right)\,e^{-\eta_1\tau^2/4}+\dots\,,
\eea
where we defined
\be\label{4.3.L1L2}
L_1(n)=\left(\frac14\ln\frac{\ep}{t}-\ln\left(2\sqrt{\tau}\,\right)\right)\delta(n+1),\qquad
L_2(n)=\ln\sqrt{1+n}\,\frac{\pa}{\pa n}.
\ee
We have to apply these formulas in \Ref{4.3.Ct1.1} to $n_i$ and to $n_{i+1}$,
\be\label{4.3.Ct1.2}
\tilde{C}_1=
\sigma\int_0^\infty\frac{d\tau}{\tau^2}\,g(2\tau^2)\left(\frac{\tau}{2}\right)^{s}
\left(\prod_{j=1}^s \int_{-1}^\infty\frac{dn_j}{\sqrt{\pi}}
\right)  \sum_{i=1}^{s-1}
\left(L_1(n_i)+L_2(n_i)\right)\left(L_1(n_{i+1})+L_2(n_{i+1})\right)
  {e^{-\eta_1\tau^2/4}} \,.
\ee
Multiplying out the two brackets,
\be\label{4.3.LL}
\left(L_1(n_i)+L_2(n_i)\right)\left(L_1(n_{i+1})+L_2(n_{i+1})\right)
=
L_1(n_{i})L_1(n_{i+1})
+2L_1(n_{i})L_2(n_{i+1})
+L_2(n_{i})L_2(n_{i+1})
\ee
(we used the symmetry under $i\to s-1-i$ in \Ref{4.3.Ct1.2}), we split $\tilde{C}_1$ into three parts,
\be\label{4.3.Ct1.3}
\tilde{C}_1=\tilde{C}_{1A}+\tilde{C}_{1B}+\tilde{C}_{1C}
\ee
and consider them separately.

We start with $\tilde{C}_{1A}$ and interchanging the orders of integration it is
\be\label{4.3.Ct1A.1}
\tilde{C}_{1A}=\sigma \pi^{-s/2}
\left(\prod_{j=1\atop j\ne i,i+1}^s \int_{-1}^\infty {dn_j}  \right)  \sum_{i=1}^{s-1}
\int_0^\infty\frac{d\tau}{\tau^3}\,g(2\tau^2)\left(\frac{\tau}{2}\right)^{s}
\left(\frac14\ln\frac{\ep}{t}-\ln\left(2\sqrt{\tau}\,\right)\right)^2\,e^{-\eta_3 \tau^2/4}
\ee
with
\be\label{4.3.eta3}
\eta_3={\eta_1}_{|n_i=n_{i+1}=-1}\,.
\ee
This is the place where the logarithm of $\ep$ appears squared.
Again, the integration over $\tau$ can be carried out explicitly
delivering lengthy formulas.
Expression \Ref{4.3.Ct1A.1} must be inserted into the energy
\Ref{4.3.E1}. Introducing the corresponding notation it is
\be\label{4.3.4.3.Ct1A.2} \Delta E_{{C}1A}=
-\frac{2}{4\pi}\sum_{s=2}^\infty \frac{\ep^2}{s+1}\,\sigma \pi^{-s/2}P(s)\,,
\ee
where
\be\label{4.3.Ps}
\int_0^\infty dt\,e^{-2t(s+1)}\tilde{C}_{1A}=\sigma\pi^{-s/2}P(s)
\ee
collects the integrations over the $n_j$ and over $t$. Again, the
$n$-integrations must be done  numerically, we were able to go up
to $s=7$. As a result we get
\be\label{4.3.Ct1A.3}
\Delta E_{{C}1A}=
\frac{R}{d^2}\, \ep\left(-8.8\,10^{-7} - 2.4\,10^{-7} \ln\ep -
3.6\,10^{-7} (\ln\ep)^2 \right).
\ee

Next we have to consider $\tilde{C}_{1B}$. In parallel to \Ref{4.3.Ct1A.1} it is
\be\label{4.3.Ct1B.1}
\tilde{C}_{1B}=
-4\sigma \pi^{-s/2}
\left(\prod_{j=1\atop j\ne i}^s \int_{-1}^\infty {dn_j}  \right)  \sum_{i=1}^{s-1}
\int_0^\infty\frac{d\tau}{\tau^3}\,g(2\tau^2)\left(\frac{\tau}{2}\right)^{s+2}
\left(\frac14\ln\frac{\ep}{t}-\ln\left(2\sqrt{\tau}\,\right)\right)
\left(2n_{i+1}-n_i-n_{i+2}\right)
\,e^{-\eta_2 \tau^2/4}
\ee
with
\be\label{4.3.eta2}
\eta_2={\eta_1}_{|n_i=-1}.
\ee
Again, the integration over $\tau$ can be carried out explicitly
and the integration over the $n_j$ only numerically. Here we could
go until $s=5$.  The result is
\be\label{4.3.Ct1B.3}
\Delta E_{{C}1B}=
\frac{R}{d^2}\, \ep
\left( -9.4\,10^{-6} + 0.000019  \ln\ep\right).
\ee

Finally we come to $\tilde{C}_{1C}$. In parallel to \Ref{4.3.Ct1A.1} it is
\bea\label{4.3.Ct1C.1}
\tilde{C}_{1C}&=&
\frac{\sigma}{2} \pi^{-s/2}
\left(\prod_{j=1 }^s \int_{-1}^\infty {dn_j}  \right)  \sum_{i=1}^{s-1}
\int_0^\infty\frac{d\tau}{\tau^3}\,g(2\tau^2)\left(\frac{\tau}{2}\right)^{s+2}
\ln(1+n_i)\ln(1+n_{i+1})
\nn\\&&    \times
\left(1+\frac{\tau^2}{2}(2n_i-n_{i-1}-n_{i+1})(2n_{i+1}-n_i-n_{i+2})\right)
\,e^{-\eta_1 \tau^2/4}\,.
\eea
As before, the $\tau$-integration can be done explicitly and the
$n$-integrations numerically. Here we went up to $s=5$. The result
is
\be\label{4.3.Ct1C.3} \Delta E_{{C}1C}= \frac{R}{d^2}\, \ep
\left( -0.000076\right). \ee
%
%%%%%%%%%%%%%%%%%%%%%%%%%%%%%%%%%%%%%%%%%%%%%%%%%%%%%%%%%%%%%%%%%%%%%%%%%%%
%%%%%%%%%%%%%%%%%%%%%%%%%%%%%%%%%%%%%%%%%%%%%%%%%%%%%%%%%%%%%%%%%%%%%%%%%%%
%%%%%%%%%%%%%%%%%%%%%%%%%%%%%%%%%%%%%%%%%%%%%%%%%%%%%%%%%%%%%%%%%%%%%%%%%%%
\subsection{The $\tilde{\Lambda}$-contribution}
We start from the formula \Ref{4.LAt} and insert
$\tilde{\lambda}_{n,n'}$ into $\cal Z^{\rm as}$, Eq.\Ref{4.Z2}.
From this we define with \Ref{3.E1} the corresponding contribution
to the energy,
\be\label{4.4.E1}
\Delta E_{\tilde{\Lambda}}=\frac{-2R}{4\pi d^2}
\sum_{s=0}^\infty\, \frac{\ep}{s+1}\int_0^\infty dt \, t\, e^{-2t(s+1)}\,\tilde{L}
\ee
with
\be\label{4.4.Lt}
\tilde{L}=\int_0^1\frac{d\tau\sqrt{\tau}}{\sqrt{1-\tau^2}}
\int_{-\infty}^\infty \frac{d\mu}{\sqrt{\pi}}\,e^{-\mu^2(s+1)/\tau}
\left(\prod_{j=1 }^s \int_{n_0}^\infty \frac{dn_j}{\sqrt{\pi}}  \right)  \sum_{0<i<j<s}
\frac{4\tau t(1-\tau^2)(1-\ep)^2\mu^2}{\gamma(n_{i})\gamma(n_{i+1})\gamma(n_{j})\gamma(n_{j+1})}
\,e^{-\eta_1}\,.\ee
Integration over $\mu$ delivers
\be\label{4.4.Lt1}
\tilde{L}=
\frac{1}{(s+1)^{3/2}}
\int_0^1\frac{d\tau \,{\tau}}{\sqrt{1-\tau^2}}
\left(\prod_{j=1 }^s \int_{n_0}^\infty \frac{dn_j}{\sqrt{\pi}}  \right)  \sum_{0<i<j<s}
\frac{2\tau^2 t(1-\tau^2)(1-\ep)^2}{\gamma(n_{i})\gamma(n_{i+1})\gamma(n_{j})\gamma(n_{j+1})}
\,e^{-\eta_1}\,.
\ee
Now we can put $\ep=0$ where it goes with $n_j$ and in $n_0$,
again keeping   all integrations and   summations finite. With
this, we note $\gamma(n)=\sqrt{\tau \,t}\sqrt{\tau+\ep/t}+\dots\,$.
As a consequence, the $n_j$-integrations become simple and are
reduced to formula \Ref{66}. Also the dependence on $i$ disappears
and the sum is $\sum_{0<i<j<s}=s(s+1)/2$. In $\tilde{L}$ only one
integration remains,
\be\label{4.4.Lt2}
\tilde{L}=\frac{s}{t(s+1)}\int_0^1d\tau\,\frac{\tau\sqrt{1-\tau^2}}{\left(\tau+\frac{\ep}{t}\right)^2}
\equiv\frac{s}{t(s+1)}\,h\left(\frac{t}{\ep}\right)\,.
\ee
This expression must be inserted into the energy \Ref{4.4.E1}.
Taking into account $h(t)=\ln(2t)-2+\dots$ for $t\to\infty$ we can
carry out the $t$-integration,
\be\label{4.4.E2}
\Delta E_{\tilde{\Lambda}}=\frac{-2R}{4\pi d^2}\,\ep
\sum_{s=0}^\infty\,\frac{s}{(s+1)^3}\left(-\frac12\ln\ep-1-\frac{\gamma}{2}-\frac12\ln(s+1)\right)+\dots\,.
\ee
Finally we carry out the summation and come to
\be\label{4.4.E3}
\Delta E_{\tilde{\Lambda}}=\frac{R}{4\pi d^2}\left[
(\zeta(2)-\zeta(3))\,(2+\gamma)
+\zeta'(3)-\zeta'(2)
+
\frac{1}{4\pi}\left(\zeta(2)-\zeta(3)\right)\, \ln\ep \right] \,\ep.
\ee
%

%
%%%%%%%%%%%%%%%%%%%%%%%%%%%%%%%%%%%%%%%%%%%%%%%%%%%%%%%%%%%%%%%%%%%%%%%%%%%
%%%%%%%%%%%%%%%%%%%%%%%%%%%%%%%%%%%%%%%%%%%%%%%%%%%%%%%%%%%%%%%%%%%%%%%%%%%%
%%%%%%%%%%%%%%%%%%%%%%%%%%%%%%%%%%%%%%%%%%%%%%%%%%%%%%%%%%%%%%%%%%%%%%%%%%%%
%%%%%%%%%%%%%%%%%%%%%%%%%%%%%%%%%%%%%%%%%%%%%%%%%%%%%%%%%%%%%%%%%%%%%%%%%%%%
\section{Conclusions}
In the forgoing section we calculated separately the different
parts contributing to the correction beyond PFA in the
electromagnetic case. These are the 'scalar' ones
(from the first two lines in \Ref{3.E4}),
\be\label{5.sc}
\frac{\Delta E_{\rm TE}+\Delta E_{\rm TM}}{\ep E_{\rm PFA}}=
\frac{1}{3}-\frac{5}{\pi^2}\approx -0.173\,,
\ee
that from parts the A, \Ref{4.1.EA1}, and B, \Ref{4.2.EB},
 in the Lambda-contribution and that from the
$\tilde{\Lambda}$ contribution, \Ref{4.4.E3},
\be\label{5.ABLt} \frac{ \Delta E_{\rm A}+  \Delta E_{\rm B}+
\Delta E_{\tilde{\Lambda}}  } {\ep E_{\rm PFA}   }=
\frac{180}{\pi^4}\left[(1+2\ln2)\zeta(3)-2(1+\ln2)\zeta(2)\right]
\approx-4.99\,, \ee
and that from part C in the Lambda-contribution, Eqs.
\Ref{4.3.Ect0_1},
\Ref{4.3.Ct1A.3},
\Ref{4.3.Ct1B.3},
\Ref{4.3.Ct1C.3},
\be\label{5.C}
\frac{\Delta E_{\tilde{C}_0}+\Delta E_{{C}1A}+\Delta E_{{C}1B}+\Delta E_{{C}1C}}{\ep E_{\rm PFA}}=
-0.045  - 0.0044  \ln\ep + 8.5 \ 10^{-6} (\ln\ep)^2\,.
\ee
In \Ref{5.ABLt} cancellations happened, for instance the
logarithmic contributions compensated  each other. This
contribution could be calculated analytically, like the 'scalar' one, Eq. \Ref{5.sc}.
The remaining
contribution \Ref{5.C} could be calculated only numerically. It
contains the remaining logarithm and  it is numerically small.
This smallness justifies the use of the quite small precision reached in subsection C.
Putting all together we come to the relative correction for the electromagnetic case,
\be\label{5.ED}
\frac{ \Delta E^{\rm ED}  } {\ep E_{\rm PFA}   }=
-5.2 - 0.0044 \ln\ep + 8.5~10^{-6} (\ln\ep)^2\,,
\ee
from which representation \Ref{eplnep} follows.
The main contribution is a constant, the logarithmic contributions are numerically
very small and do not play a role at any reasonable separation.

The result \Ref{5.ED} is quite unexpected since it it quite large.
It must be mentioned that it is not in agreement with the
numerical calculations performed in
\cite{Emig:2008zz,MaiaNeto:2008zz}, where $\sim 1.4$ for the
constant contribution was obtained. In these calculations the fits
were made without accounting for possible logarithmic
contributions. However, in view of the smallness of the
coefficients in \Ref{5.ED} this should be acceptable.

The result \Ref{5.ED} also does not support the  experimental
results found in \cite{Decca2007}. However, the experiments were
done with real metals. It should be a subject of future work to
account for that in an analytical calculation.

In general, it must be underlined that \Ref{5.ED} is the second
term in an asymptotic expansion. Therefore it is hard to predict
how small $\ep$ must be to get a good approximation in this way.
In principle, it cannot be excluded that \Ref{5.ED} gives a good
approximation only for $\ep$ smaller than that which are
interesting for the experiments and which are accessible by the
mentioned numerical approaches.

There is still another problem with the first correction beyond
PFA for Neumann boundary conditions. Already in the easier case of
a cylinder in front of a plane the numerical calculations reported
in \cite{Lombardo:2008ww} showed good agreement with the
analytical ones in \cite{Bordag:2006vc} only for Dirichlet
boundary conditions, but not for Neumann ones. For the latter,
only with a fit including a logarithmic term in the second order,
$\ep^2\ln\ep$, agreement was found. For a sphere in front of a
plane for a scalar field, the numerical results reported in
\cite{Emig:2008zz} are in agreement with the analytical ones,
Eq.\Ref{3.E4}, only for Dirichlet conditions on both, the sphere
and the plane (DD). In the other three combinations of boundary
conditions the numbers are quite different. This is quite
unexpected for the case of Neumann conditions on the plane but
Dirichlet conditions on the sphere (ND), since the analytical
results for (DD) and (ND) are the same.

It should be mentioned that the appearance of logarithms in the
expansion is probably a rather common feature. This can be seen
from the structure of the corrections, see, for example, Eqn.(B13)
in \cite{Bordag:2006vc}. The expansion parameter $\ep$ is always
accompanied by a factor $1/t$, producing in higher orders a
singularity at $t\to0$. It is only in the order $\ep$ considered
in \cite{Bordag:2006vc} as well as in \cite{BORDAG2008C} that
these did not show up.

In view of the agreement of the numerical results with the
analytical results in the (DD) case,  the disagreement in the
other three cases where we have at least on one surface Neumann
conditions can be viewed as a hint that the numerical approach is
more difficult once Neumann conditions are involved. It could
happen that an agreement can be reached for smaller $\ep$ only. We
would like to point out that also in the analytical approach the
calculations with Neumann conditions are a bit more delicate. The
point is in the convergence of the sum over $s$ in \Ref{3.E1}.
While for Dirichlet conditions the decrease is $\sim(s+1)^{-4}$
(first line in \Ref{3.E3}), it is only $\sim(s+1)^{-2}$ for
Neumann conditions. In the electromagnetic case it is even weaker,
$\sim(s+1)^{-2}\ln(s+1)$, for example in \Ref{4.4.E2}.

As a consequence of the mentioned disagreement it would be
interesting to improve the numerical approach. The main obstacle
is that very large orbital momenta must be accounted for. A way
out could consist of three steps. First, one may   expand the
logarithm as in Eq.\Ref{2.E3}. As we know from the analytical
approach this sum is converging. To get a satisfactory precision,
to take a few terms should be sufficient. In the second step one
would make the substitution \Ref{3.subst} also in the numerical
approach. This allows to capture the main contribution. Finally,
as third step, one would need to adopt some approach of coarsening
to the orbital momentum summations or their substitution by
integrals. In any case, further work is necessary in this
direction.

\vspace{1cm}\noindent
V.N. was supported by the Swedish Research Council
(Vetenskapsr{\aa}det),  grant 621-2006-3046.
The authors benefited from exchange of ideas by the ESF Research Network CASIMIR.

\end{document}